\RequirePackage{lineno}
\documentclass[prd,nofootinbib,twocolumn,superscriptaddress,preprintnumbers,
balancelastpage,longbibliography]{revtex4-1}

\usepackage{graphicx}
\usepackage{color}
\usepackage{hepunits}
\usepackage{amsmath}
\usepackage{booktabs}
\usepackage{array}
\usepackage{makecell}
\usepackage[framemethod=TikZ]{mdframed}
\usepackage[colorlinks=true,
urlcolor=blue,
anchorcolor=blue,
citecolor=blue,
filecolor=blue,
linkcolor=red,
menucolor=blue,
linktocpage=true,
pdfproducer=medialab,
pdfa=true
]{hyperref}
\usepackage[utf8]{inputenc}
\usepackage[english]{babel}

\newcolumntype{C}[1]{>{\centering\let\newline\\\arraybackslash\hspace{0pt}}m{#1}}

\newcommand{\be}{\begin{equation}}
\newcommand{\ee}{\end{equation}}
\newcommand{\bea}{\begin{equation}\begin{aligned}}
\newcommand{\eea}{\end{aligned}\end{equation}}

\begin{document}

\title{A Search for Neutrino Point-Source Populations in 7 Years \\
of IceCube Data with Neutrino-count Statistics}

\affiliation{III. Physikalisches Institut, RWTH Aachen University, D-52056 Aachen, Germany}
\affiliation{Department of Physics, University of Adelaide, Adelaide, 5005, Australia}
\affiliation{Dept. of Physics and Astronomy, University of Alaska Anchorage, 3211 Providence Dr., Anchorage, AK 99508, USA}
\affiliation{Dept. of Physics, University of Texas at Arlington, 502 Yates St., Science Hall Rm 108, Box 19059, Arlington, TX 76019, USA}
\affiliation{CTSPS, Clark-Atlanta University, Atlanta, GA 30314, USA}
\affiliation{School of Physics and Center for Relativistic Astrophysics, Georgia Institute of Technology, Atlanta, GA 30332, USA}
\affiliation{Dept. of Physics, Southern University, Baton Rouge, LA 70813, USA}
\affiliation{Dept. of Physics, University of California, Berkeley, CA 94720, USA}
\affiliation{Lawrence Berkeley National Laboratory, Berkeley, CA 94720, USA}
\affiliation{Institut f{\"u}r Physik, Humboldt-Universit{\"a}t zu Berlin, D-12489 Berlin, Germany}
\affiliation{Fakult{\"a}t f{\"u}r Physik {\&} Astronomie, Ruhr-Universit{\"a}t Bochum, D-44780 Bochum, Germany}
\affiliation{Universit{\'e} Libre de Bruxelles, Science Faculty CP230, B-1050 Brussels, Belgium}
\affiliation{Vrije Universiteit Brussel (VUB), Dienst ELEM, B-1050 Brussels, Belgium}
\affiliation{Dept. of Physics, Massachusetts Institute of Technology, Cambridge, MA 02139, USA}
\affiliation{Institute for Data, Systems, and Society, Massachusetts Institute of Technology, Cambridge, MA 02139, USA}
\affiliation{Dept. of Physics and Institute for Global Prominent Research, Chiba University, Chiba 263-8522, Japan}
\affiliation{Dept. of Physics and Astronomy, University of Canterbury, Private Bag 4800, Christchurch, New Zealand}
\affiliation{Dept. of Physics, University of Maryland, College Park, MD 20742, USA}
\affiliation{Dept. of Astronomy, Ohio State University, Columbus, OH 43210, USA}
\affiliation{Dept. of Physics and Center for Cosmology and Astro-Particle Physics, Ohio State University, Columbus, OH 43210, USA}
\affiliation{Niels Bohr Institute, University of Copenhagen, DK-2100 Copenhagen, Denmark}
\affiliation{Dept. of Physics, TU Dortmund University, D-44221 Dortmund, Germany}
\affiliation{Dept. of Physics and Astronomy, Michigan State University, East Lansing, MI 48824, USA}
\affiliation{Dept. of Physics, University of Alberta, Edmonton, Alberta, Canada T6G 2E1}
\affiliation{Erlangen Centre for Astroparticle Physics, Friedrich-Alexander-Universit{\"a}t Erlangen-N{\"u}rnberg, D-91058 Erlangen, Germany}
\affiliation{Physik-department, Technische Universit{\"a}t M{\"u}nchen, D-85748 Garching, Germany}
\affiliation{D{\'e}partement de physique nucl{\'e}aire et corpusculaire, Universit{\'e} de Gen{\`e}ve, CH-1211 Gen{\`e}ve, Switzerland}
\affiliation{Dept. of Physics and Astronomy, University of Gent, B-9000 Gent, Belgium}
\affiliation{Dept. of Physics and Astronomy, University of California, Irvine, CA 92697, USA}
\affiliation{Karlsruhe Institute of Technology, Institut f{\"u}r Kernphysik, D-76021 Karlsruhe, Germany}
\affiliation{Dept. of Physics and Astronomy, University of Kansas, Lawrence, KS 66045, USA}
\affiliation{SNOLAB, 1039 Regional Road 24, Creighton Mine 9, Lively, ON, Canada P3Y 1N2}
\affiliation{Department of Physics and Astronomy, UCLA, Los Angeles, CA 90095, USA}
\affiliation{Department of Physics, Mercer University, Macon, GA 31207-0001}
\affiliation{Dept. of Astronomy, University of Wisconsin, Madison, WI 53706, USA}
\affiliation{Dept. of Physics and Wisconsin IceCube Particle Astrophysics Center, University of Wisconsin, Madison, WI 53706, USA}
\affiliation{Institute of Physics, University of Mainz, Staudinger Weg 7, D-55099 Mainz, Germany}
\affiliation{Department of Physics, Marquette University, Milwaukee, WI, 53201, USA}
\affiliation{Institut f{\"u}r Kernphysik, Westf{\"a}lische Wilhelms-Universit{\"a}t M{\"u}nster, D-48149 M{\"u}nster, Germany}
\affiliation{Bartol Research Institute and Dept. of Physics and Astronomy, University of Delaware, Newark, DE 19716, USA}
\affiliation{Dept. of Physics, Yale University, New Haven, CT 06520, USA}
\affiliation{Dept. of Physics, University of Oxford, Parks Road, Oxford OX1 3PU, UK}
\affiliation{Dept. of Physics, Drexel University, 3141 Chestnut Street, Philadelphia, PA 19104, USA}
\affiliation{Physics Department, South Dakota School of Mines and Technology, Rapid City, SD 57701, USA}
\affiliation{Dept. of Physics, University of Wisconsin, River Falls, WI 54022, USA}
\affiliation{Dept. of Physics and Astronomy, University of Rochester, Rochester, NY 14627, USA}
\affiliation{Oskar Klein Centre and Dept. of Physics, Stockholm University, SE-10691 Stockholm, Sweden}
\affiliation{Dept. of Physics and Astronomy, Stony Brook University, Stony Brook, NY 11794-3800, USA}
\affiliation{Dept. of Physics, Sungkyunkwan University, Suwon 16419, Korea}
\affiliation{Dept. of Physics and Astronomy, University of Alabama, Tuscaloosa, AL 35487, USA}
\affiliation{Dept. of Astronomy and Astrophysics, Pennsylvania State University, University Park, PA 16802, USA}
\affiliation{Dept. of Physics, Pennsylvania State University, University Park, PA 16802, USA}
\affiliation{Dept. of Physics and Astronomy, Uppsala University, Box 516, S-75120 Uppsala, Sweden}
\affiliation{Dept. of Physics, University of Wuppertal, D-42119 Wuppertal, Germany}
\affiliation{DESY, D-15738 Zeuthen, Germany}

\author{M. G. Aartsen}
\affiliation{Dept. of Physics and Astronomy, University of Canterbury, Private Bag 4800, Christchurch, New Zealand}
\author{M. Ackermann}
\affiliation{DESY, D-15738 Zeuthen, Germany}
\author{J. Adams}
\affiliation{Dept. of Physics and Astronomy, University of Canterbury, Private Bag 4800, Christchurch, New Zealand}
\author{J. A. Aguilar}
\affiliation{Universit{\'e} Libre de Bruxelles, Science Faculty CP230, B-1050 Brussels, Belgium}
\author{M. Ahlers}
\affiliation{Niels Bohr Institute, University of Copenhagen, DK-2100 Copenhagen, Denmark}
\author{M. Ahrens}
\affiliation{Oskar Klein Centre and Dept. of Physics, Stockholm University, SE-10691 Stockholm, Sweden}
\author{C. Alispach}
\affiliation{D{\'e}partement de physique nucl{\'e}aire et corpusculaire, Universit{\'e} de Gen{\`e}ve, CH-1211 Gen{\`e}ve, Switzerland}
\author{K. Andeen}
\affiliation{Department of Physics, Marquette University, Milwaukee, WI, 53201, USA}
\author{T. Anderson}
\affiliation{Dept. of Physics, Pennsylvania State University, University Park, PA 16802, USA}
\author{I. Ansseau}
\affiliation{Universit{\'e} Libre de Bruxelles, Science Faculty CP230, B-1050 Brussels, Belgium}
% ...

\collaboration{IceCube Collaboration}
\noaffiliation

\preprint{LCTP-19-19}

\begin{abstract}
The presence of a population of point sources in a dataset modifies the underlying neutrino-count statistics from the Poisson distribution.
This deviation can be exactly quantified using the non-Poissonian template fitting technique, and in this work we present the first application this approach to the IceCube high-energy neutrino dataset.
Using this method, we search in 7 years of IceCube data for point-source populations correlated with the disk of the Milky Way, the \textit{Fermi} bubbles, the Schlegel, Finkbeiner, and Davis dust map, or with the isotropic extragalactic sky.
No evidence for such a population is found in the data using this technique, and in the absence of a signal we establish constraints on population models with source count distribution functions that can be described by a power-law with a single break.
The derived limits can be interpreted in the context of many possible source classes.
In order to enhance the flexibility of the results, we publish the full posterior from our analysis, which can be used to establish limits on specific population models that would contribute to the observed IceCube neutrino flux.
\end{abstract}

\maketitle

%%%%%%%%%%%%%%%%%%%%%%%%%%%%%%%
\section{Introduction}\label{sec:Introduction}
%%%%%%%%%%%%%%%%%%%%%%%%%%%%%%%

The conclusive discovery of an astrophysical neutrino flux at IceCube~\cite{Aartsen:2013bka,Aartsen:2013jdh,Aartsen:2014gkd,Aartsen:2014muf,Aartsen:2015knd,Aartsen:2015rwa} presents a new window through which we can view the Universe.
With energies in the PeV range, these neutrinos free stream to Earth over scales where extragalactic photons of the same energy are attenuated.
This makes the IceCube neutrino window novel not only in terms of messenger, but also as it provides insight into extreme energy phenomena.
While recently evidence was found for a point source of neutrinos~\cite{IceCube:2018dnn}, at present the origin of a large fraction of astrophysical flux remains unknown.
For a review see~\cite{Spiering:2017kkh}.

With an observed flux close to the Waxman-Bahcall bound~\cite{Waxman:1998yy}, the leading hypothesis posits that the IceCube neutrinos are produced through extragalactic hadronic processes, where high-energy proton interactions produce charged pions, which in turn decay to produce neutrinos.\footnote{The neutrino flavor ratio observed with IceCube is presently consistent with a pionic origin~\cite{Aartsen:2015ivb}.}
There are a number of viable models for the origin of these hadronic collisions involving conventional astrophysical sources, see~\cite{Murase:2016gly} for a comprehensive discussion.
Nevertheless some of the promising source classes already appear to be disfavored as the sole origin of the observed cosmic neutrinos.
The current limits on the presence of point sources in the data, which we will soon discuss, place under tension a pure blazar origin for the observed flux, although they may still contribute~\cite{Padovani:2014bha,Padovani:2015mba,Petropoulou:2015upa,Murase:2016gly} -- especially in light of the recent discovery~\cite{IceCube:2018dnn}.
Further, there are claims that a dominant starburst galaxy origin is in tension with existing gamma-ray measurements~\cite{Tamborra:2014xia,Bechtol:2015uqb,Sudoh:2018ana}.
Viable source classes remain, however, such as radio galaxies~\cite{Murase:2016gly,Hooper:2016jls,Blanco:2017bgl}.
Regardless of their origin, due to the hadronic origin of the neutrinos in these scenarios, a definitive identification of neutrinos from a particular source class would provide a deep insight into the wider problem of high-energy cosmic-ray acceleration~\cite{Learned:2000sw,Halzen:2002pg,Anchordoqui:2009nf,Anchordoqui:2013dnh}.
As such the implications of the IceCube dataset for various source classes, even those coming from null results, are leading to important insights into high-energy astrophysics as a whole.

As long as the question of their origin remains unanswered, however, there will be room for speculation as to a potentially more exotic origin for these neutrinos.
A possibility that has received significant attention in the literature is that these neutrinos could be produced in the decay of $\sim$PeV scale dark matter, see, for example~\cite{Feldstein:2013kka,Esmaili:2013gha,Ema:2013nda,Bhattacharya:2014vwa,Zavala:2014dla,Chianese:2016smc,Chianese:2017nwe,Schmaltz:2017oov}.
Whilst extragalactic dark matter decays would be distributed isotropically, decays within the Milky Way would imprint a directional anisotropy within the data in such a scenario.
Although the present data appears to be isotropic~\cite{Aartsen:2015knd}, the measurements have not yet resulted in enough significance to disfavor the dark matter scenarios~\cite{Esmaili:2014rma}.
More stringent constraints arise as generically such models produce photons or charged cosmic-rays in addition to the neutrinos, and limits on dark matter from such final states tend to disfavor many scenarios, see~\cite{Kalashev:2016cre,Cohen:2016uyg,Kuznetsov:2016fjt,Kalashev:2017ijd,Abeysekara:2017jxs}.
Nevertheless the possibility remains that there could be hints as to the origin of dark matter within the IceCube dataset, a possibility which highlights the importance of a definitive determination of the neutrino origins.

Due to their lack of electric charge, neutrinos remain unattenuated as they travel through the magnetic fields that permeate the universe.
The neutrinos point back to their source of origin, raising the possibility that if the incident neutrino direction can be measured accurately enough, the source class could be identified.
A single event is unlikely to be determinative, but statistical analyses applied to larger datasets can search for clustering on the sky, which can indicate the presence of point sources.
A number of such searches for point sources have already been performed within IceCube~\cite{Abbasi:2010rd,Aartsen:2013uuv,Aartsen:2013dla,Aartsen:2014cva,Aartsen:2014ivk,Aartsen:2015yva,Aartsen:2016tpb,Aartsen:2016oji,Glauch:2017xwr,Aartsen:2018ywr}, but have not yet found evidence of such clustering.
The absence of any statistically significant clustering in these analyses thus far suggests that whatever is the primary contributor to the IceCube flux is not a small number of bright sources, but rather a larger population of sources.
In the present work we extend this line of investigation through the application of a novel technique for searching for populations of sources, which should be viewed as complementary to existing and ongoing individual source searches.
We accomplish this using a technique known as the Non-Poissonian Template Fit (NPTF), which has found widespread application to the \textit{Fermi} LAT gamma-ray data, but is applied here to the IceCube neutrino data for the first time.

The basic principle underpinning the NPTF is that in the presence of unmodeled point sources, the neutrino count statistics of a dataset is distinct with respect to a Poisson distribution.
The requirement that the point sources are unmodeled is central to the method; indeed, if a point source is resolved and has a known location, a model for it can be constructed and the observed data will represent a Poisson draw from that model.
The NPTF, however, remains agnostic as to the location of the sources and simply accounts for the fact that a population of point sources will result in larger upward and downward fluctuation than can be produced by the Poisson distribution.
As we will review in this work, that deviation can be rigorously quantified into a likelihood, which can test the preference for non-Poissonian statistics in the data, and thereby uncover evidence for point-source population.

The NPTF has a number of advantages over traditional point source search techniques.
The method is naturally couched in the language of populations of sources, as the fundamental object constrained is the source-count distribution $\mathrm{d} N/\mathrm{d} F$, that is the distribution of sources with a flux between $F$ and $F+\mathrm{d} F$.
By way of contrast, the standard techniques search for sources of a given flux $F'$ one at a time.
These methods are often calibrated against simulations where $N$ point sources each of flux $F'$ have been injected, which corresponds to the special case $\mathrm{d}N/\mathrm{d}F = N \delta(F-F')$.
Studies using a similar method to the NPTF applied to the public High Energy Starting Events (HESE) data have shown that population based approaches can, under certain assumptions, probe deeper than traditional methods and search for sources too dim to be resolved individually~\cite{Feyereisen:2016fzb,Feyereisen:2017fnk}.
Such techniques have also be used to extend limits on the brightest possible source within the data~\cite{Ando:2017xcb}.
Furthermore, as the NPTF is a method for template fitting, it can readily incorporate a non-trivial spatial dependence for the sources.
Template fitting is a technique of fitting data with models following a predetermined spatial distribution, or templates.
We will exploit this to search for sources correlated with the disk of the Milky Way and the \textit{Fermi} bubbles, in addition to extragalactic sources which are distributed isotropically.
Even in the case of isotropically distributed extragalactic sources, due to the spatial variation inherent in the IceCube effective area matrix, the resulting neutrino count map will be non-isotropic.
This complication can be readily handled in the NPTF.
On the other hand, there are drawbacks to the method.
The NPTF is fundamentally a binned technique, which precludes optimization through the use of event by event reconstruction information.
Moreover, in this work the NPTF is restricted to be used in a single energy bin.
This is not a fundamental limitation of the method, but simply a shortcoming in existing implementations of it.
Yet, at present this implies that we are unable to account for the strong variation of the response to neutrinos in the IceCube detector as a function of energy, other than through optimizing the choice of energy bin.
Taken together these highlight the complementary nature of the NPTF based search presented in this work to alternative approaches to the problem.
We emphasize that the NPTF is distinct from other novel searches for point sources that have been applied to the IceCube dataset, in that it searches for the specific modification to the neutrino-count statistics imprinted by a point source population.
An example of such techniques is the multipole and two-point autocorrelation approach~\cite{Aartsen:2014ivk,Glauch:2017xwr}, which search for a statistical increase in the number of events with small angular separation that point sources would induce.
Fundamentally, point sources produce additional clustering within the data on the scale of the instrument point spread function, which is usually smaller than the scale on which the diffuse backgrounds cluster.
In that sense, the NPTF is also looking to exploit similar information, but it approaches the problem by instead considering how a population of point sources modifies the statistics of the neutrino count map.
In addition to its different approach, the NPTF also offers a number of advantages over these techniques.
Most importantly, being a template method, it can readily account for the spatial variation expected for galactic sources, or non-isotropic detector response, as discussed above.
Further, the NPTF can be formulated in terms of an analytic likelihood, as we will review in this work, and this allows for an efficient practical implementation of the method.
The NPTF is also distinct to the search for steady point sources with specific flux-characteristics, as considered in~\cite{Aartsen:2018ywr}, where a test-statistic is estimated for a given population model with parameterized source density and luminosity. For the NPTF, a broader range of population models are tested, allowing the shape of the source distribution to be additionally parameterized.
Constraints on models with specified flux-characteristics can, however, be derived from the NPTF results, and we will demonstrate this explicitly by placing constraints on the space of standard candle luminosity functions using \texttt{FIRESONG}~\cite{Taboada:2018gmk}, similar to those considered in~\cite{Aartsen:2018ywr}.

The remainder of the discussion will be structured as follows.
In Sec.~\ref{sec:EventSelection} we outline the event selection used to distill the dataset analyzed in this work.
Section~\ref{sec:NPTF} is dedicated to a review of the NPTF method, and a detailed description of the challenges in applying the method at IceCube, and the associated solutions.
Following from this, in Sec.~\ref{sec:ExpSens} we determine the expected sensitivity of the method based on Monte Carlo simulations.
Then, in Sec.~\ref{sec:Results} we show the result of applying the NPTF to the real IceCube data, and in the absence of a signal derive constraints.
The full posterior from our analysis is made publicly available,\footnote{\url{https://icecube.wisc.edu/science/data/NPTF_7yr_posterior}} with the details of the file discussed in App.~\ref{sec:PubPost}.
Finally our conclusions are presented in Sec.~\ref{sec:Conclusion}.

%%%%%%%%%%%%%%%%%%%%%%%%%%%%%%%
\section{Event Selection}\label{sec:EventSelection}
%%%%%%%%%%%%%%%%%%%%%%%%%%%%%%%

%%%%%%%%%%%%%%%%%%%%%%%%%%%%%%%
\subsection{The IceCube Neutrino Observatory}
%%%%%%%%%%%%%%%%%%%%%%%%%%%%%%%

IceCube is a cubic-kilometer Cherenkov detector, which is composed of 5160 digital optical modules (DOMs) embedded in the Antarctic ice at the South Pole~\cite{Achterberg:2006md}.
These DOMs are attached to 86 strings of cable at depths between 1450-2450 m beneath the surface of the ice.
Most of the DOMs have a vertical spacing of 17 m along the strings and the average distance between neighboring strings is $\sim$125 m.
Each module consists of a photomultiplier tube, onboard digitization board, and a separate board with LEDs for calibration~\cite{Abbasi:2008aa, Abbasi:2010vc}.

Construction of IceCube started in 2004 and was completed in December 2010.
Before the full detector was completed, data was being taken in partial configurations with fewer than 86 strings.
In the present work we make use of 7 years of IceCube data.
The first three years were taken during the 40-string (IC-40), 59-string (IC-59), and 79-string (IC-79) configurations, as described in Refs.~\cite{Aartsen:2013uuv,Abbasi:2010rd}.
The subsequent four years of data exploited the full 86-string (IC-86) configuration, as outlined in Refs.~\cite{Aartsen:2014cva,Aartsen:2016oji}.

%%%%%%%%%%%%%%%%%%%%%%%%%%%%%%%
\subsection{Neutrino Detection at IceCube}
%%%%%%%%%%%%%%%%%%%%%%%%%%%%%%%

Neutrinos are notoriously difficult to detect, and just because they point back to their origin does not mean we can necessarily extract that direction.
As we cannot detect the neutrinos directly, the challenge is in inferring the direction from visible products left behind from a neutral or charged current interaction the neutrino undergoes within or in the vicinity of IceCube.
In order to enhance our sensitivity, we choose to focus on events where the flight direction of the neutrino can be accurately reconstructed, as explained below.

In detail, there are three different event topologies within IceCube produced by neutrino interactions: tracks, cascades, and double-bangs.
Tracks result from muons traversing the detector, while cascades result from the charged-current interactions of electron or tau neutrinos or neutral-current interactions of any neutrino.
The interactions in cascade events produce an almost spherical light emission making directional reconstruction difficult.
Tracks from muons of $\sim$TeV or greater energies can travel several kilometers while constantly emitting Cherenkov light, making them ideal candidates for an accurate directional reconstruction.
Double-bangs result from charged current interactions of tau neutrinos at very high energies where the tau lepton decays to hadrons far enough from the initial interaction to create two distinct cascades, and the first candidates for such events have now been identified~\cite{Stachurska:2018}.
Since tracks are the optimal topology for directional accuracy, in this paper we will only use tracks.

The only neutrino interactions that can create tracks at TeV energies are charged-current interactions of muon neutrinos (and muon antineutrinos).
Track events can be further divided into two subclasses: starting tracks occur when a muon neutrino has its charged-current interaction inside of the detector volume, while through-going tracks occur when that interaction occurs outside of the detector volume.
Through-going tracks can result from any high-energy muon, including muons produced by cosmic-ray interactions in the atmosphere, but they also have a much higher effective area for muon neutrinos than starting tracks since the charged-current interaction can occur in a much larger volume than the detector volume.
In this paper we will consider only events reconstructed as through-going tracks, in order to take advantage of the high number of events resulting from the much larger effective area, however the lower purity of astrophysical neutrinos in the sample creates a background we will need to account for in the NPTF analysis.

%%%%%%%%%%%%%%%%%%%%%%%%%%%%%%%
\subsection{Data and Simulations}
%%%%%%%%%%%%%%%%%%%%%%%%%%%%%%%

With these motivations, the specific dataset used in this analysis was the through-going tracks in IceCube's 7 year point source sample~\cite{Aartsen:2016oji}, and we refer to Sec.~2.2 of that reference for a detailed account of the dataset.
The 7 year point source sample has been accrued through several different event selections.
Events from the year of IC-40 data were selected using fixed selection criteria on several parameters~\cite{Abbasi:2010rd}, while events from the remaining years~\cite{Aartsen:2013uuv, Aartsen:2014cva, Aartsen:2016oji} were selected using multivariate boosted decision trees (BDTs) to classify events as signal or background.
The BDTs were trained with separate background and signal datasets.
In the background case the BDT was trained on the data itself, a procedure which is justified as the data is known to be strongly dominated by the background.
For the signal, the BDT was instead trained on muon neutrino Monte Carlo simulations. 
These same Monte Carlo simulated muon neutrinos are also used to calculate the effective area of the detector and the point spread function (PSF).
More information on the Monte Carlo simulation can be found in Ref.~\cite{Aartsen:2016xlq}.
Through these selection processes the sample is divided into two regions, up-going (with declination $\delta > -5^{\circ}$ ) and down-going ($\delta < -5^{\circ}$).
The down-going region is dominated by atmospheric muons, while the up-going region is shielded from these muons by the Earth.
The up-going region is dominated by atmospheric neutrinos.
Despite this distinction, in our analysis we will not distinguish between up and down-going events, instead performing a full sky analysis.
As in Ref.~\cite{Aartsen:2016oji}, the total livetime is 2,431 days, with 422,791 up-going events, and 289,078 down going.
The directions of the events in this sample were reconstructed using a likelihood-based method, which uses information on how the photons scatter and get absorbed with the ice~\cite{Ahrens:2003fg}.
The reconstructions of events in the IC-86 data used a more advanced description of the ice~\cite{Aartsen:2013rt} and a better parameterization of how the photons interact with it~\cite{Aartsen:2014cva}. 
The muon energy is estimated by approximating its energy loss along its reconstructed track~\cite{Aartsen:2013vja}. 
In order to fine tune this sample for our analysis we make a cut on the reconstructed muon energy.
The expected energy spectrum for an astrophysical neutrino flux is harder than the energy spectrum for the atmospheric neutrino flux.
It should also be noted that the reconstructed muon energy is only an estimate of the muon energy at a point where it enters the detector and thus can only provide an estimated lower-bound of the primary neutrino energy, as much of the energy of the event can be deposited outside the detector.
Nevertheless, a cut on the reconstructed muon energy can still effectively increase the sample's purity with respect to the harder-spectrum astrophysical flux.
We determine the optimal energy cut to be 10$^{0.5}$ TeV $\approx 3$ TeV.
This value is determined by maximizing the sensitivity for the full sky analysis presented in this work.
We note that as the energy distribution of the background in the Northern and Southern hemispheres is quite different, if we performed an analysis restricting to either of these the optimal energy cut would vary.
We also make a spatial cut around the poles of the detector ($|\delta| > 85^{\circ}$), as the scrambling procedure is less effective, and the reconstructions can be poorly behaved, in these regions.
These cuts lead us to our final sample of 309,134 events.

Since the NPTF is a binned analysis we must spatially bin the data.
For this purpose we use \texttt{HEALPix}~\cite{Gorski:2004by} skymaps to bin the data into pixels of equal solid angle.
There is still a freedom in terms of how large to choose these bins, controlled by the \texttt{nside} parameter, however we find that a value of 64 maximizes our full sky sensitivity, which corresponds to 49,152 bins, each of size approximately 0.84 square degrees.

%%%%%%%%%%%%%%%%%%%%%%%%%%%%%%%
\section{The Non-Poissonian Template Fit}\label{sec:NPTF}
%%%%%%%%%%%%%%%%%%%%%%%%%%%%%%%

The NPTF quantifies the following observation into a rigorous analytic likelihood: compared to a count map following a Poisson distribution, a map determined by a distribution of sources will have more hot and cold pixels -- the hot pixels associated with locations where there are sources, and the cold pixels where there are none.
In slightly more detail, there are at least two steps involved in going from an underlying point source distribution to a neutrino count map.
Firstly, we need to determine how many point sources are expected and how they are distributed on the sky.
Secondly, given a $\mathrm{d}N/\mathrm{d}F$, we must determine the map of neutrino counts that is expected from this distribution of sources.
The NPTF likelihood provides the rigorous answer to these questions, as well as incorporating additional complications arising through detector effects such as the finite PSF of the instrument.
This likelihood, when applied to the data, allows for a determination as to whether such a population of sources is preferred, and if not constraints on $\mathrm{d}N/\mathrm{d}F$ can be set.

Although the NPTF is a relatively recent method, core aspects have long been employed in astronomy.
The fundamental observation that point source distributions lead to more hot and cold pixels is at the heart of the $P(D)$ method that has been applied to X-ray datasets~\cite{1993A&A...275....1H,1993MNRAS.262..619G,Miyaji:2001dp,Gendreau:1997di,Perri:2000fv}.
The method was extended to gamma-rays in~\cite{Malyshev:2011zi}, where the analytic likelihood was also first derived.
The likelihood was extended to the NPTF, i.e. a full template based method, in~\cite{Lee:2015fea}, and the NPTF or similar methods have found a number of additional applications in gamma-ray astronomy~\cite{Lee:2014mza,Feyereisen:2015cea,Zechlin:2015wdz,Linden:2016rcf,Zechlin:2016pme,Lisanti:2016jub,DiMauro:2017ing}.
As mentioned above, a related method that is similarly population based has been considered in the context of publicly available IceCube data~\cite{Feyereisen:2016fzb,Feyereisen:2017fnk}.
The NPTF has now been incorporated into a publicly available code \texttt{NPTFit}~\cite{Mishra-Sharma:2016gis},\footnote{\url{https://github.com/bsafdi/NPTFit/}} which we use for the present work.

The remainder of this section will begin with a more quantitative review of the method, outlining the core ideas leading to the NPTF likelihood.
After this, we focus on some of the challenges that had to be addressed in order to apply this method to the IceCube dataset, which in particular required a careful treatment of the instrument effective area matrix and PSF.
Thirdly, we will describe how we can combine the NPTF likelihood with additional Poissonian models that we will use in our hypothesis testing, and finally, we outline our inference procedure which we will use to search for point-source populations in the data.

%%%%%%%%%%%%%%%%%%%%%%%%%%%%%%%
\subsection{Overview of the Method}
%%%%%%%%%%%%%%%%%%%%%%%%%%%%%%%

In this section we provide a brief, quantitative review of the NPTF likelihood framework, particularly emphasizing aspects that will be relevant to an application at IceCube.
A more comprehensive description of the method and a derivation of all quoted expressions can be found in Ref.~\cite{Mishra-Sharma:2016gis}.

Our ultimate goal is to write down a likelihood for a set of model parameters ${\boldsymbol \theta}$, given the data $d$ described in Sec.~\ref{sec:EventSelection} -- i.e.,\ we want a function $\mathcal{L}(d \vert {\boldsymbol \theta})$.
Let us start with a description of the model parameters.
In the case where we only have neutrinos originating from a single point-source population, then the model parameters specify the source-count distribution $\mathrm{d}N/\mathrm{d}F$.
In principle it is possible to keep the form of $\mathrm{d}N/\mathrm{d}F$ very general, but a particularly simple analytic expression for the source-count function that is likely a good approximation to many realistic neutrino source classes is a broken power-law\footnote{In passing we note that \texttt{NPTFit} can handle a broken power-law source-count function with an arbitrary number of breaks.}
\be
\frac{\mathrm{d}N_p}{\mathrm{d}F} (F; {\boldsymbol \theta}) = A\,T_p 
\left\{ 
\begin{array}{lc} 
(F/F_b)^{-n_1} & F \geq F_b\,, \\
(F/F_b)^{-n_2} & F < F_b\,.
\end{array}
\right.
\label{eq:dNdF}
\ee
The use of a common functional form to describe the flux of both galactic and extragalactic sources, is motivated by the fact that this ansatz appears to be a reasonable description of both populations in gamma-rays observed by \textit{Fermi}, see for example~\cite{Lee:2015fea}.
In Eq.~\ref{eq:dNdF} we have added the term $T_p$ to the source count function.
$T_p$ is a template, or pixelated spatial map describing the spatial distribution of the sources on the sky.
It is the only term with an explicit pixel-by-pixel variation.
As an example, for an isotropic or extragalactic distribution of sources, we have $T_p = 1$ for every value of $p$.
This template could be considered as a model parameter and fitted for, but in our analysis we will take $T_p$ to be fixed before performing any likelihood analysis.
With the template fixed, there are four model parameters for a singly broken power-law: the normalization $A$, location of the break $F_b$, and power-law indices above and below the break $n_1$ and $n_2$.
Thus formally ${\boldsymbol \theta} = \{A, F_b, n_1, n_2\}$.

To provide some intuition for these parameters, note that we can calculate the total expected number of point sources and also the total expected flux from the population in each pixel from direct integration of the source-count function over all possible fluxes as follows
\bea
N_p^{\rm PS} = &\int_0^{\infty} \mathrm{d}F\,\frac{\mathrm{d}N_p}{\mathrm{d}F} = \frac{A\,T_p\,F_b(n_1-n_2)}{(n_1-1)(1-n_2)}\,, \\
F_p^{\rm PS} = &\int_0^{\infty} \mathrm{d}F\,F\frac{\mathrm{d}N_p}{\mathrm{d}F} = \frac{A\,T_p\,F_b^2 (n_1-n_2)}{(n_1-2)(2-n_2)}\,.
\label{eq:TotalNF}
\eea
In performing these integrals, we require $n_1 > 1$ and $n_2 < 1$ to obtain a finite $N_p^{\rm PS}$, whereas for a finite $F_p^{\rm PS}$ we need $n_1 > 2$ and $n_2 < 2$.
There is an important distinction between the number of sources and flux as given in~Eq.~\ref{eq:TotalNF}.
The total expected flux per pixel, $F_p^{\rm PS}$, is related to the expected number of neutrinos observed through the IceCube effective area, and in this sense is tied to an observable in the data.
Yet the total expected number of sources, $N_p^{\rm PS}$, is not tied to an observable derivable from a map of neutrino counts.
In this sense, a real map could have a best fit value $n_2 \in [1,2]$, which corresponds to an infinite number of sources but a finite flux.
While in practice the total number of sources should be finite, the effective number of sources, for realistic source populations such as star-forming galaxies~\cite{Tamborra:2014xia}, can appear infinite.
This effect occurs because the cut-off in the source-count distribution, $\mathrm{d}N/\mathrm{d}S$, that makes the number of sources finite, appears at flux values well below the level where the sources contribute, on average, more than one photon or neutrino~\cite{Lisanti:2016jub}.

When presenting results it will be helpful to use a different set of variables instead of $\{A, F_b, n_1, n_2\}$.
Specifically we will replace $A$ and $F_b$ with the expected number of point sources across the whole sky, $N^{\rm PS}$, and the expected flux per source, $\bar{F}^{\rm PS}$.
The change of variables can be implemented as follows,
\bea
N^{\rm PS} &= \sum_p N_p^{\rm PS} = \frac{A F_b (n_1 - n_2)}{(n_1 - 1)(1 - n_2)} \sum_p T_p\,, \\
\bar{F}^{\rm PS} &= \frac{\sum_p F_p^{\rm PS}}{\sum_p N_p^{\rm PS}} = F_b \frac{(n_1-1)(1-n_2)}{(n_1-2)(2-n_2)}\,.
\label{eq:NbarF}
\eea
Given a template and value of $n_1$ and $n_2$, we can then change variables to these more intuitive quantities, which we will exploit when presenting our results.

At this point we comment on an important assumption that will be used throughout our analysis.
We have repeatedly discussed the flux of sources, $F$, eschewing the question of what energy this flux is being measured at.
To resolve this, we will assume that our point-source population follows the canonical astrophysical expectation of $E^{-2}$.
In detail the flux from an individual source is given by
\be
\frac{\mathrm{d} \Phi(E)}{\mathrm{d} E} = F \left( \frac{E}{1~{\rm TeV}} \right)^{-2}\,,
\label{eq:PhiE}
\ee
which serves to contextualize the quantity $F$ discussed throughout this section.
It should be thought of as the normalization constant for the energy dependent flux, and we will assume an identical $E^{-2}$ scaling for all sources.
If the actual spectrum is softer than this, the events will be shifted to lower energies where the backgrounds are higher, and generically we would expect a reduction of the sensitivities shown here.

Returning to our derivation of the likelihood, the source-count function parametrizes our model prediction for the population of sources, but our goal is to embed this into a likelihood that can be fit to neutrino count data.
As a first step towards this we need to address the fact that the discussion so far has been couched in the language of fluxes per neutrino source, $\Phi$, commonly quoted with units of [neutrinos/cm$^2$/s], whereas what is observed in the instrument is an integer number of neutrino events.
The conversion between these two variables is provided by the effective area matrix, which accounts for the fact that two point sources with equal flux at different locations on the sky will contribute a different number of detected neutrinos within IceCube.
The effective area is the amalgamation of the detection efficiency for neutrinos incident on the IceCube detector from different directions, as well as an accounting for the fact the detector has a fixed location at the South Pole.

The conversion from flux, $F$, in units of [neutrinos/cm$^2$/s] to counts, $S$, is achieved with the combination of the effective area and collection time, usually called an exposure map, which we denote by $\mathcal{E}_p$ -- a pixel dependent quantity due to the spatial dependence in the effective area matrix.
We defer the discussion of how the appropriate $\mathcal{E}_p$ map for our analysis was derived until the following subsection.
Assuming for now the map is known, then using this we can then convert to a source-count function in terms of counts rather than flux as follows,
\be
\frac{\mathrm{d}N_p}{\mathrm{d}S}(S; {\boldsymbol \theta}) = \frac{1}{\mathcal{E}_p} \frac{\mathrm{d}N_p}{\mathrm{d}F}(F = S/\mathcal{E}_p; {\boldsymbol \theta})\,.
\label{eq:dNdS}
\ee
As $\mathcal{E}_p$ can take on a different value in every single pixel, the conversion from flux to counts should in truth be performed in every pixel.
Nevertheless this is in practice often unnecessary.
It is usually sufficient to divide the full map into a number of subregions, which each have comparable $\mathcal{E}_p$ values, take the mean $\mathcal{E}_p$ in this region, and perform the conversion once per region.
Within the \texttt{NPTFit} framework, the number of subregions is controlled by the keyword \texttt{nexp}, and results are commonly convergent for small values of this parameter.
As the appropriate $\mathcal{E}_p$ for our dataset varied significantly, albeit smoothly, as a function of declination, we chose to use 50 exposure regions in order to ensure the transformation from flux to counts was accurately performed.

From $\mathrm{d}N_p/\mathrm{d}S$, we can then move towards the NPTF likelihood by deriving the following useful quantity: the expected number of sources that will contribute $m$ neutrinos in a pixel $p$, $x_{p,m}({\boldsymbol \theta})$.
To do so, note that $\mathrm{d}N_p/\mathrm{d}S$ evaluated at a particular $S$ provides the expected number of sources that contribute an expected number of counts $S$, where of course $S$ does not need to be an integer.
The probability that one such source provides $m$ neutrinos is then determined by the Poisson distribution, specifically $S^m e^{-S}/m!$.
From here, $x_{p,m}({\boldsymbol \theta})$ is given by weighting this factor by the source-count distribution, and integrating over all $S$, as each value could Poisson fluctuate to $m$.
In detail
\be
x_{p,m}({\boldsymbol \theta}) = \int_0^{\infty} \mathrm{d}S\, \frac{\mathrm{d} N_p}{\mathrm{d} S} (S; {\boldsymbol \theta} )\frac{S^m e^{-S}}{m!}\,.
\label{eq:xpm}
\ee
This expression, whilst intuitive, has an inherent assumption that will be invalidated in most real applications: if a point source is located in pixel $p$, then it deposits all of its observed neutrinos in that same pixel.
This neglects the finite PSF at IceCube, but we will hold off on a discussion of how to address this until the next subsection.

Our final goal of this subsection is to move from $x_{p,m}$ to the probability of observing $k$ neutrinos in a pixel $p$, $p_{p,k}$.
Combining $p_{p,k}$ with the observed number of neutrinos in the data, $d_p$, and then taking the product over all pixels $p$, exactly gives us the likelihood through which we can constrain $\mathrm{d}N/\mathrm{d}F$, or more specifically ${\boldsymbol \theta} = \{A, F_b, n_1, n_2\}$.
Being fully explicit, we have:
\be
\mathcal{L}\left( d \vert {\boldsymbol \theta} \right) = \prod_{p} p_{p,k=d_p}({\boldsymbol \theta})\,,
\label{eq:likelihood}
\ee
where the product is taken over all pixels in the dataset analyzed.
In order to derive an expression for $p_{p,k}$, we use the concept of probability generating functions.\footnote{For a review, see for example Section 3.6 of \cite{Bardin:1999ak}.}
If we have a discrete probability distribution described by a set $\left\{ p_{p,k} \right\}$ known for all $k$, then the generating function in a given pixel is defined as
\be
P_p(t) \equiv \sum_{k=0}^{\infty} p_{p,k} t^k\,,
\label{eq:GenFn}
\ee
where $t$ is an auxiliary variable.
The probabilities can be recovered from $P_p(t)$ via
\be
p_{p,k} = \left. \frac{1}{k!} \frac{\mathrm{d}^k P_p(t)}{\mathrm{d}t^k} \right|_{t=0}\,.
\label{eq:inversion}
\ee
In the case of models described by the Poisson distribution with expected counts $\mu_p$, substituting the Poisson distribution into Eq.~\ref{eq:GenFn}, reveals the associated generating function to be $P_p(t) = \exp [ \mu_p (t-1)]$.

Next, the aim is to construct the associated non-Poissonian generating function, starting with the expected number of $m$-neutrino sources, $x_{p,m}({\boldsymbol \theta})$, as given in Eq.~\ref{eq:xpm}.
Of course $m$ can take on any integer value, but for the moment let us take it to be fixed, and we will determine the generating function for $m$-neutrino sources, denoted $P_p^{(m)}(t)$.
From the definition in Eq.~\ref{eq:GenFn}, we need to know the probability of seeing $k$ neutrinos in the pixel $p$, given by $p_{p,k}$, a value that will depend on how many $m$-neutrino sources there are.
Specifically, $k$ must be some integer $n_m$ multiple of $m$, where $n_m$ drawn from a Poisson distribution with mean $x_{p,m}({\boldsymbol \theta})$,
\be
p_{p,n_m} = \frac{x_{p,m}^{n_m} e^{-x_{p,m}}}{n_m!}\,,
\ee
where we have left the dependence on the parameters ${\boldsymbol \theta}$ implicit.
In terms of this, the probability for seeing $k$ neutrinos in pixel $p$ is simply $p_{p,n_m}$ for the case that $k = n_m \times m$, or zero otherwise as we are still keeping $m$ fixed.
Substituting this information into Eq.~\ref{eq:GenFn}, we obtain
\bea
P_p^{(m)}(t) &= \sum_{n_m} t^{m\,n_m}\,\frac{x_{p,m}^{n_m} e^{-x_{p,m}}}{n_m!} \\
&= \exp \left[ x_{p,m} (t^m - 1) \right]\,,
\eea
where we only included the non-zero values in the sum.
Now this result was obtained for a fixed $m$, in order to obtain the full non-Poissonian generating function, we need to account for all possible $m$.
As each source is independent, each value of $m$ is independent also.
We can then make use of the fact that the generating function of a sum of independent random variables is given by the product of each variable's generating function.
Accordingly, the full non-Poissonian generating function is given by
\be
P_p(t) = \exp \left[ \sum_{m=1}^{\infty} x_{p,m} \left( t^m - 1 \right) \right]\,.
\label{eq:GenNP}
\ee
From here, using the inversion formula in~Eq.~\ref{eq:inversion}, the probability of observing $k$ neutrinos in a pixel $p$ is
\be
p_{p,k} = \left. \frac{1}{k!} \frac{\mathrm{d}^k}{\mathrm{d}t^k} \exp \left[ \sum_{m=1}^{\infty} x_{p,m} \left( t^m - 1 \right) \right] \right|_{t=0}\,,
\ee
or combining this with our earlier results
\begin{align}
p_{p,k}({\boldsymbol \theta}) = &\frac{1}{k!} \frac{\mathrm{d}^k}{\mathrm{d}t^k} \exp \left[ \sum_{m=1}^{\infty} \left( t^m - 1 \right)
\right. \\
&\left. \left.\int_0^{\infty} \mathrm{d}S\, \frac{1}{\mathcal{E}_p} \frac{\mathrm{d} N_p}{\mathrm{d} F} \left(\frac{S}{\mathcal{E}_p}; {\boldsymbol \theta} \right)\frac{S^m e^{-S}}{m!}
\right] \right|_{t=0}\,. \nonumber
\end{align}
This expression, combined with~Eq.~\ref{eq:likelihood}, gives us our full NPTF likelihood as a function of the source-count function $\mathrm{d}N_p/\mathrm{d}F$, which is exactly what we want to constrain.
Although this expression contains an unevaluated integral, in the case of a multiply broken power-law it can be efficiently implemented as the integral can be calculated analytically, and furthermore the $p_{p,k}$ can be evaluated recursively in $k$.
All these details are described in Ref.~\cite{Mishra-Sharma:2016gis}.

Below we will outline how this likelihood can be extended to account for the presence of additional contributions to the data beyond just point sources.
But before doing so we turn to the practicalities of implementing the NPTF in the specific case of IceCube.

%%%%%%%%%%%%%%%%%%%%%%%%%%%%%%%
\subsection{Implementation at IceCube}
%%%%%%%%%%%%%%%%%%%%%%%%%%%%%%%

There are two immediate obstacles to implementing the NPTF at IceCube, beyond the basic outline described in the previous subsection.
Firstly, we need to calculate an appropriate effective area matrix in order to determine $\mathcal{E}_p$, which appears in the conversion between the $\mathrm{d}N/\mathrm{d}F$ and $\mathrm{d}N/\mathrm{d}S$, as per~Eq.~\ref{eq:dNdS}.
Secondly, we must incorporate the real IceCube PSF and thereby remove the assumption hidden in~Eq.~\ref{eq:xpm}, that a source deposits all of its neutrinos in the pixel it is located.
For some of the neutrinos recorded at IceCube, there is a small but non-zero probability that their true incident direction is separated from the reconstructed value by a significant (angular) distance, so this effect must be accounted for in essentially any binning of the data.
We will address each point in turn.

Consider first the effective area matrix, a quantity which specifies the response of the detectors to an incident neutrino of a given energy, right ascension, and declination.
This, of course, cannot be calculated analytically.
Instead we take simulations of individual events and use these to construct $\mathcal{E}_p$.
We do this by reweighting simulated events within our energy range according to an $E^{-2}$ spectrum, as this is the spectrum that we assume our point-source population follows, see~Eq.~\ref{eq:PhiE} and the surrounding discussion.
This then provides the average detector response as a function of right ascension and declination, which we can map to galactic coordinates and provide $\mathcal{E}_p$.
As claimed above, this map can vary significantly, by more than an order of magnitude between locations with highest and lowest effective area.
Given IceCube's location at the geographic South Pole, this variation is exclusively in declination, which tracks whether the event is arriving above from through the atmosphere, or below through the Earth.
In light of this we used a relatively large number of 50 exposure regions to convert from flux to counts.

Turning next to the PSF, recall as discussed already that the NPTF likelihood derived in the last subsection assumed perfect angular reconstruction of every event.
This assumption was invoked in writing down the number of sources contributing $m$ neutrinos in a pixel $p$, denoted $x_{p,m}$, in~Eq.~\ref{eq:xpm}.
By moving directly from the expected number of sources in the pixel to the expected number of neutrino counts, implicit is the assumption that the source deposits all of its flux into that pixel.
Yet detector effects will smear the flux of a real source amongst a number of pixels.
As in the NPTF we do not keep track of which pixels are adjacent, what we want is the distribution for how a given source deposits its flux amongst the pixels on the map, a quantity denoted $\rho(f)$.
Here $f \in [0,1]$ is the fraction of the point source's flux; the case of near perfect angular reconstruction, as compared to the pixel size, corresponds to a $\rho(f)$ peaked near $f=1$, as most of the flux tends to be distributed in one pixel (the pixel where the source is located).
Indeed in the limit of exact angular reconstruction, we have $\rho(f) = 2\delta(f-1)$.\footnote{Note that in this idealized case, $\rho(f)$ will also have a contribution at $f=0$ as for perfect angular reconstruction most of the sky will receive no flux.
Nevertheless, when $f=0$ there is no contribution to the neutrino flux, and so in practice we will always neglect the zero flux case.}
More generically, however, imperfect angular reconstruction leads to a distribution peaked nearer $f=0$ as most often a pixel will only get a small fraction of the flux.
As a concrete example, consider a $3 \times 3$ grid with a point source at the center.
Imagine the source deposits 60\% of its flux in the central pixel that it inhabits, and then 5\% in each of the 8 pixels surrounding it.
In this case, many more pixels experience a small amount of flux, and so $\rho(f)$ would still be peaked towards smaller values of $f$.
Further note that $\rho(f)$ itself is not a probability density function, instead as the point source must deposit all of its flux somewhere, the distribution is normalized so that
\be
\int_0^1\mathrm{d}f\,f\,\rho(f) = 1\,.
\label{eq:frhof}
\ee

Imagine we have the appropriate $\rho(f)$ -- we will describe how to derive this shortly -- consider how this modifies $x_{p,m}$.
Previously, we used the fact that $\mathrm{d}N_p/\mathrm{d}S$ provides the number of sources that contribute an expected number of counts $S$, to then reweight this quantity by the probability of fluctuating from $S$ to $m$, given by the Poisson distribution $S^m e^{-S}/m!$.
Now, however, a source is only expected to deposit a fraction $f$ of that flux in the pixel under consideration, and so instead the reweighting to obtain $m$ neutrinos is $(fS)^m e^{-fS}/m!$.
Further the probability that a given value of $f$ is chosen is dictated by the distribution $\rho(f)$, and integrating over all possible flux fractions we arrive at the following modification for $x_{p,m}$:
\bea
x_{p,m}({\boldsymbol \theta}) = &\int_0^{\infty} \mathrm{d}S\, \frac{\mathrm{d} N_p}{\mathrm{d} S} (S; {\boldsymbol \theta} ) \\
\times &\int_0^1 \mathrm{d}f\, \rho(f) \frac{(f S)^m e^{-f S}}{m!}\,.
\label{eq:xpmPSF}
\eea
Another way to understand the above modification is to compare this result to~Eq.~\ref{eq:xpm}.
Doing so, the modification induced by the finite PSF is seen to be equivalent to substituting in a modified source-count function,
\be
\frac{\mathrm{d} N_p}{\mathrm{d} S} (S) \to \frac{\mathrm{d} \tilde{N}_p}{\mathrm{d} S} (S) = \int_0^1 \mathrm{d} f\, \frac{\rho(f)}{f} \frac{\mathrm{d} N_p}{\mathrm{d} S} (S/f)\,.
\label{eq:modsourcecount}
\ee
Propagating the modification in~Eq.~\ref{eq:xpmPSF} through to the NPTF likelihood then gives a full accounting for the effect of the finite PSF, once we have $\rho(f)$.
Note that taking $\rho(f) = 2\delta(f-1)$, i.e. perfect angular reconstruction, the above expression reduces to~Eq.~\ref{eq:xpm}, as it must, and this can also be seen clearly in~Eq.~\ref{eq:modsourcecount}.

All that remains then in order to incorporate the IceCube PSF is an algorithm for determining $\rho(f)$.
Commonly, the instrument PSF is stated as a probability distribution for a given event to be located at some radius from the center of a source.
The median reconstruction angle for $\nu_{\mu} + \bar{\nu}_{\mu}$ events can be seen, for example, in Fig.~2 of~\cite{Aartsen:2016oji}, where the angular error can vary from half a degree to several degrees depending on the energy and event type.
Nevertheless, this is just the median reconstruction angle; the tails of this distribution are considerably non-Gaussian, and can extend out to very wide angles in the case of poorly reconstructed events.
Modeling the tails correctly is critical for an NPTF analysis.
As an example of why this is important, if there is a true population of sources all with identical fluxes, a mismodeled PSF can lead to the fit preferring an additional population of lower flux sources associated with mis-reconstructed neutrinos.

The event-by-event determination of the angular reconstruction can be exploited in an unbinned analysis, as done for example in~\cite{Aartsen:2016oji}, however as the NPTF is fundamentally a binned method we will instead consider quantities averaged within the dataset of interest.
There is no simple analytic expression for converting from the known PSF to $\rho(f)$, as, for example -- unlike the PSF -- $\rho(f)$ depends critically on the binning of the underlying map.\footnote{To highlight this note that in the limit where the map contains only one pixel, we must have $\rho(f)=2\delta(f-1)$, independent of the PSF.}
We can, however, determine this distribution using the following algorithmic prescription:
\begin{enumerate}
\item Simulate $N$ equal flux point sources on a blank map with the same pixelization as the NPTF will be applied to;
\item Determine the fraction of the total flux in each pixel, $f_p$, defined such that $\sum_p f_p = 1$;
\item Define a flux binning $\Delta f$, and as a function of flux $f$, define $\Delta n(f)$ the number of pixels that have a flux between $f$ and $f + \Delta f$; and
\item Combine these quantities to define $\rho(f)$ as follows
\be
\rho(f) \approx \frac{\Delta n(f)}{N \Delta f}\,.
\label{eq:rhof}
\ee
\end{enumerate}
The above relation is approximate, and only becomes exact in the limit $N \to \infty$ and $\Delta f \to 0$.

\begin{figure}[t]
\begin{center}
\includegraphics[scale=0.4]{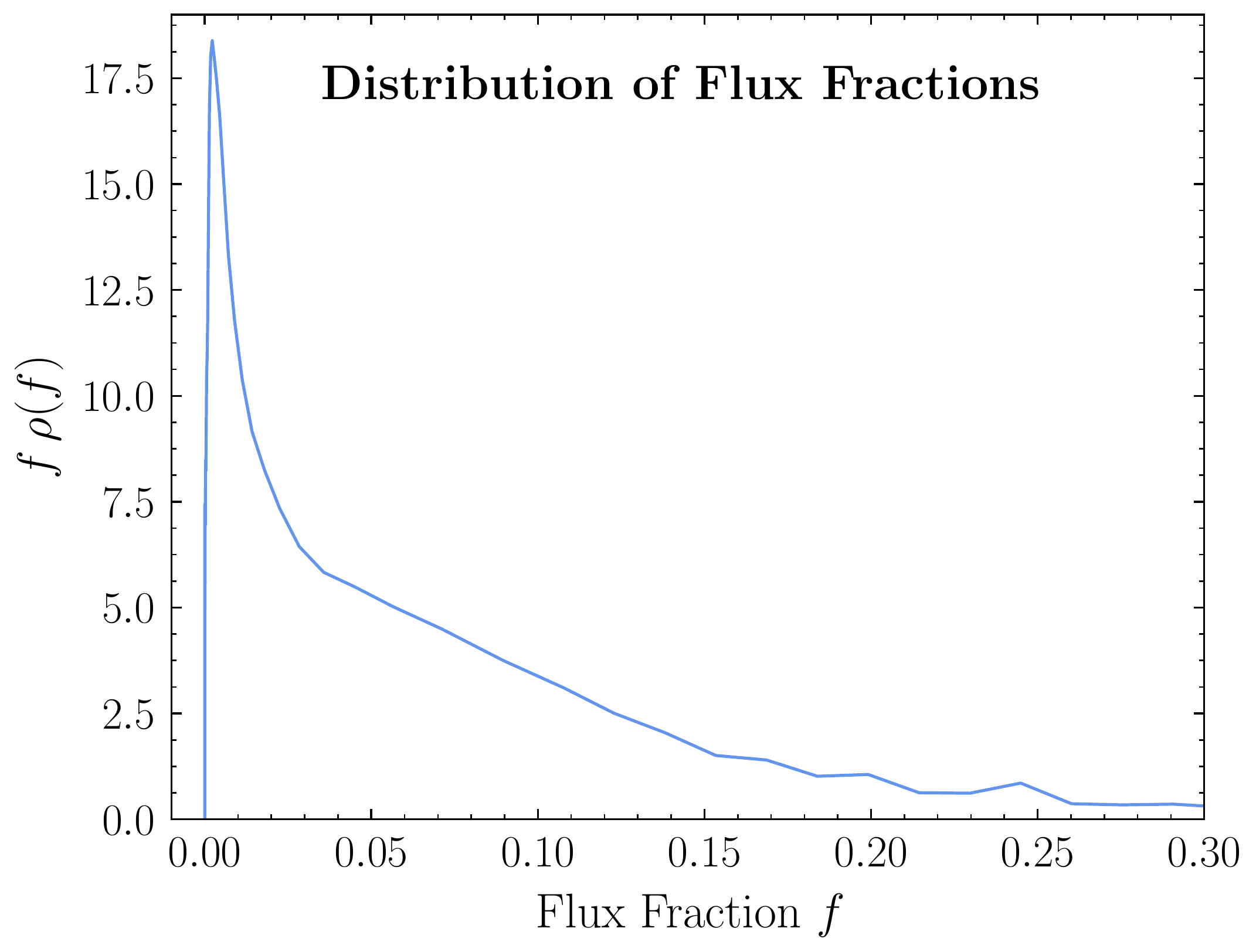}
\end{center}
\vspace{-0.2cm}
\caption{The distribution of the frequency of pixels that contain a fraction $f$ of the flux from a point source, $\rho(f)$, for the full sky template, appropriate for modeling isotropic extragalactic source.
This quantity is central to incorporating the PSF of IceCube into the NPTF likelihood, according to~Eq.~\ref{eq:xpmPSF}.
We have chosen to show $f\,\rho(f)$, as following~Eq.~\ref{eq:frhof} this quantity integrates to one given conservation of flux.
See text for details.
}
\label{fig:PSF-rhof}
\vspace{-0.4cm}
\end{figure}

In order to simulate this in practice we deposit a large number of sources on the sky, and for each one model the neutrino distribution according to the appropriate PSF at each location.
To account for the energy dependence inherent in the PSF of IceCube, following~Eq.~\ref{eq:PhiE}, we assume each source has an $E^{-2}$ spectrum, and draw events for the source according to this distribution.
In this way we can exactly build up $\rho(f)$ as defined in~Eq.~\ref{eq:rhof}, and the result is shown in Fig.~\ref{fig:PSF-rhof} for our default full-sky analysis.
In that figure we have zoomed in on the small $f$ values where the distribution is peaked.
That the flux fraction distribution is peaked at small values is indicative of the fact that the IceCube PSF has tails that extend significantly more broadly than the size of a pixel on the map, but is also quite generic of $\rho(f)$ unless the angular reconstruction is significantly better than the pixel size.
For comparison, note that the linear size of our pixels is $\sim$$0.92^{\circ}$, which is comparable to the median reconstruction angle of our events which is $\sim$$1^{\circ}$.
As the PSF of IceCube varies across the sky, $\rho(f)$ must be determined for each of our spatial distributions separately.\footnote{In principle, due to the spatial variation of the PSF, $\rho(f)$ also varies spatially.
The approximation made in this work is that we will use the mean of $\rho(f)$ across the sky, as weighted by our various spatial templates.}

In light of these results the NPTF likelihood can now be applied to the IceCube dataset.
As an important validation of the method, we performed extensive tests based upon Monte Carlo studies where we injected and then recovered point-source populations.
An example of this is shown in Fig.~\ref{fig:injsig}, discussed in detail below.
We emphasize again that in particular the details discussed in this section were critical to the successful recovery of injected populations.
With the method validated, we can now look to calculate the expected sensitivity at IceCube, which we turn to in Sec.~\ref{sec:ExpSens}.
Before doing so, however, in the next two subsections we will discuss how we incorporate backgrounds that are not associated with point sources into our model, and then how these various likelihoods will be combined into an inference framework we can use to test for the presence of point sources.

%%%%%%%%%%%%%%%%%%%%%%%%%%%%%%%
\subsection{Adding Poissonian models}
%%%%%%%%%%%%%%%%%%%%%%%%%%%%%%%

At the very least due to the presence of irreducible backgrounds, we know that point sources cannot be the only contribution to the IceCube dataset, and in this section we discuss how to augment the NPTF likelihood to account for these.
These additional contributions are generally expected to be described by the Poisson distribution, following an underlying spatial map.
To incorporate both the Poissonian statistics and the spatial variation, we will use the language of Poissonian templates, following the language from a recent application of this topic in Ref.~\cite{Lisanti:2017qoz}.

To begin with, as in the NPTF case, we imagine our model follows a spatial distribution that once pixelized can be described by a map $T_p$.
Unlike for an NPTF model, where $T_p$ specified the spatial distribution of point sources, in the Poisson case we require $T_p$ to be proportional to the expected distribution of counts, not flux.
As a concrete example, in the case where our model is for isotropic extragalactic neutrino emission, the appropriate $T_p$ would still have a spatial variation inherited from the effective area matrix described in the previous subsection.
As in the non-Poissonian case, we assume for a given model the Poissonian template, $T_p$, is specified ahead of time.
What we fit for in this case is the overall normalization of this template, in terms of which our Poissonian model prediction is given by
\be
\mu_p({\boldsymbol \theta}) = A\,T_p\,,
\ee
so that ${\boldsymbol \theta} = \{ A \}$, where we emphasize that $A$ has no pixel dependence.
Given that the sum of two Poisson distributions with means $\mu_1$ and $\mu_2$ is again a Poisson distribution of mean $\mu_1 + \mu_2$, we can readily extend this formalism to account for multiple Poisson models.
For example if we had $n$ of these, described by template $T_p^1,\ldots,T_p^n$, then our combined model prediction in each pixel would be
\be
\mu_p({\boldsymbol \theta}) = \sum_{\ell=1}^n A^{\ell} \,T_p^{\ell}\,,
\ee
where now ${\boldsymbol \theta} = \{ A^1, \ldots, A^n \}$.
To provide a concrete example, we may want to model the observed flux using a model that combines three sources: 1. terrestrial backgrounds such atmospheric neutrinos; 2. diffuse extragalactic emission; and 3. diffuse emission from the Milky Way.
In such a scenario, we would have three Poissonian templates, and for each of these $T_p^{\ell}$ would describe the pixel dependence of the flux, and $A^{\ell}$ the overall normalization.
For the case of extragalactic emission, as the flux is expected to be isotropic, $T_p^{\ell}$ would then be a map of the IceCube detector's response to a uniform incident flux.

In terms of this, if all we had was a set of Poissonian models then we could write down our likelihood according to the Poisson distribution,
\be
p_{p,k}({\boldsymbol \theta}) = \frac{\mu_p({\boldsymbol \theta})^k e^{-\mu_p({\boldsymbol \theta})}}{k!}\,.
\label{eq:Poisppk}
\ee
Nonetheless, we want to construct a likelihood incorporating both Poissonian and non-Poissonian models.
For this purpose, we can use the property of generating functions we exploited earlier, specifically that the generating function which describes the sum of two independent random variables is given by the product of the generating functions for the individual variables.
For the Poisson case, given in~Eq.~\ref{eq:Poisppk}, the associated generating function according to~Eq.~\ref{eq:GenFn}, is given by
\be
P_p(t; {\boldsymbol \theta}) = \exp \left[ \mu_p({\boldsymbol \theta}) (t-1) \right]\,.
\ee
Combined with the generating function for the non-Poissonian case given in~Eq.~\ref{eq:GenNP}, we arrive at
\bea
P_p(t; {\boldsymbol \theta}) = \exp &\left[ \sum_{m=1}^{\infty} x_{p,m}({\boldsymbol \theta}) \left( t^m - 1 \right) \right. \\
&\left.\vphantom{\sum_{m=1}^{\infty}}+\,\mu_p({\boldsymbol \theta}) (t-1) \right]\,,
\label{eq:PandNPgen}
\eea
where now ${\boldsymbol \theta} = \left\{ A, F_b, n_1, n_2, A^1, \ldots, A^n \right\}$.
Through the use of~Eq.~\ref{eq:inversion}, this generating function can be used to derive a combined likelihood that includes both Poissonian and non-Poissonian models, accomplishing one of the main aims of this subsection.

\begin{figure}[t]
\begin{center}
\includegraphics[scale=0.3]{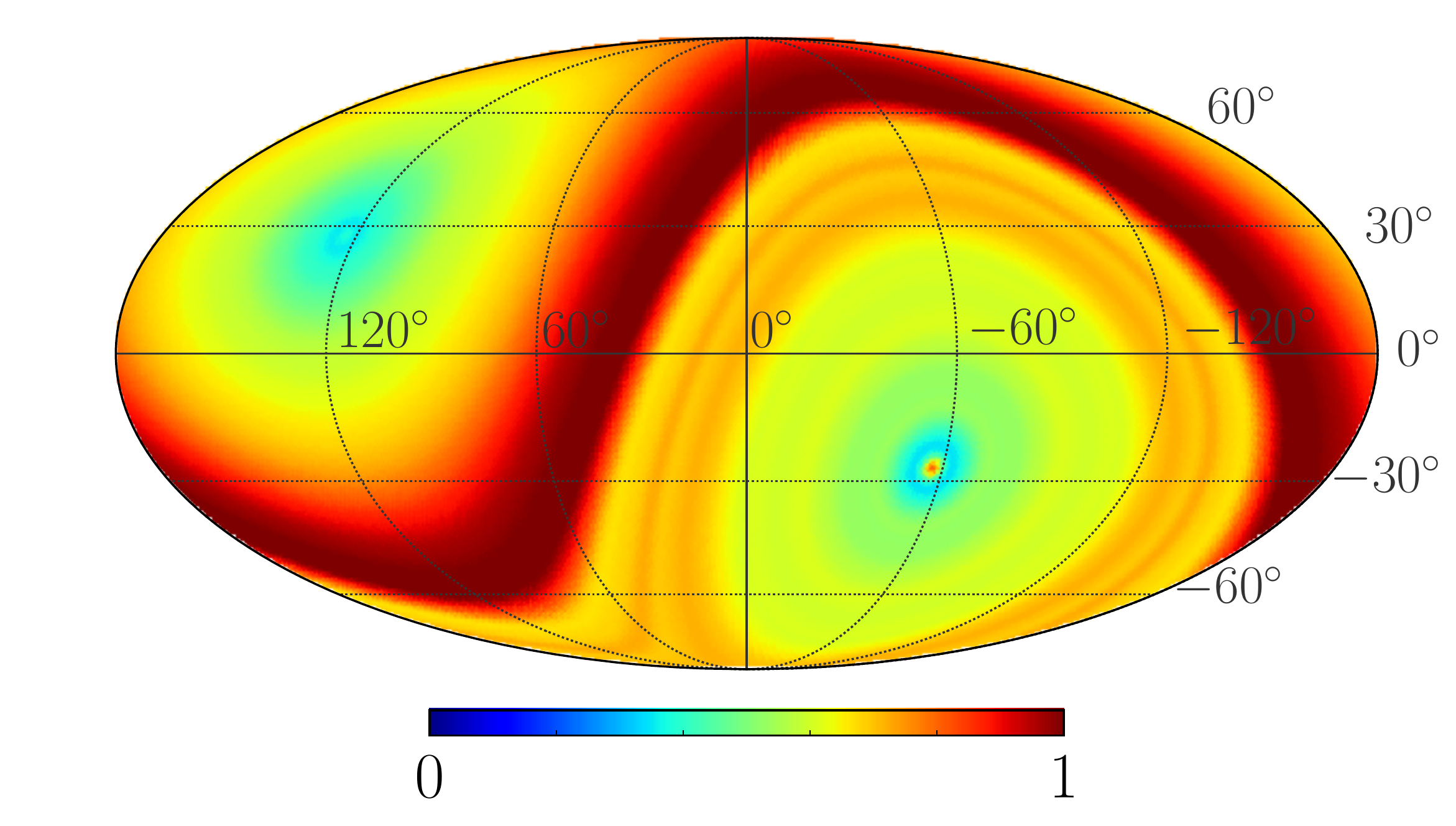}
\end{center}
\vspace{-0.5cm}
\caption{Data driven background template for the spatial distribution of atmospheric neutrinos and muons, derived using the procedure described in the text.
This map is referred to as $T_p^{\rm bkg}$.
The map is a Mollweide projection of an underlying distribution in galactic coordinates, and the overall normalization is arbitrary.
}
\label{fig:BkgMap}
\vspace{-0.4cm}
\end{figure}

The main application for the Poissonian template formalism in our work will be to model the known backgrounds arising from atmospheric neutrinos and muons.
For this purpose we need to derive an appropriate $T_p$ describing the spatial distribution of these contributions.
Determining this from first principles is at present out of reach.
Fortunately, however, we can estimate the distribution from the data alone.
The reason for this is if we assume the data is made up of predominantly background events and a subset of point sources, then we can remove the point sources in the following way.
Given the approximate azimuthal symmetry of the effective area of detector, we can take the data collected by IceCube and scramble the events by assigning them a random right ascension value.
This process removes any point source hotspots, as they will be smeared out along bands of constant declination.
Furthermore, as the background is only expected to vary with declination, this process does not degrade the spatial information pertaining to the background process.
Applying this process once gives a map that is still as noisy as the data.
In order to extract a map more appropriate for the mean of a Poisson distribution, we repeated this scrambling process a large number of times and take the average of the resulting maps.
Finally, to remove the noise in declination, we convolve this model with a von Mises–Fisher distribution that has a concentration corresponding to $1.08^{\circ}$, chosen as this is the median angular resolution at $\sim$1 TeV.
This last step can be justified as the real data has been scrambled on such a scale due to the PSF.

The map resulting from this procedure is shown in Fig.~\ref{fig:BkgMap}, which is the Mollweide projection of the map in galactic coordinates.
The most apparent feature in this map are the strong variation away from the poles towards the equator where the largest flux is observed.
Taking this map as template, we can readily incorporate the largest expected background into our likelihood.
In subsequent discussions we will refer to this map as $T_p^{\rm bkg}$.

%%%%%%%%%%%%%%%%%%%%%%%%%%%%%%%
\subsection{Inference Framework}
%%%%%%%%%%%%%%%%%%%%%%%%%%%%%%%

So far in this section we have introduced the formalism required to calculate the likelihood for a dataset in the presence of a population of sources and additional Poissonian contributions.
Our goal now is to put this machinery to work in the form of a test statistic that we can use to test for the presence of point sources in the data.
Our test statistic will compare between two hypotheses: that point sources, distributed according to a spatial template $T_p^{\rm PS}$, are present in the data or they are not.
We will refer to these as the non-Poissonian and Poissonian hypotheses respectively.

\begin{table}[t]
\def\arraystretch{1.2}
\begin{tabular}{C{4.5cm}C{1.8cm}}
\toprule
\Xhline{3\arrayrulewidth}
Parameter & Prior Range \\ \hline
$\log_{10} \big( A$ [TeV$\,$cm$^2\,$s]$\big)$ & $[2.41,16.41]$ \\
$\log_{10} \big( F_b$ [TeV$^{-1}\,$cm$^{-2}\,$s$^{-1}$]$\big)$ & $[-18,-7]$ \\
$n_1$ & $[2,10]$ \\
$n_2$ & $[-6,2]$ \\
$A^{\rm bkg}$ & $[0.5,1.5]$ \\
$A^{\rm P}$ & $[-0.5,1]$ \\
\bottomrule
\Xhline{3\arrayrulewidth}
\end{tabular}
\caption{List of priors used in calculating the marginal likelihood.
All priors are taken to be flat in the given ranges.
}
\label{tab:priors}
\end{table}

\begin{figure*}[t]
\begin{center}
\includegraphics[scale=0.245]{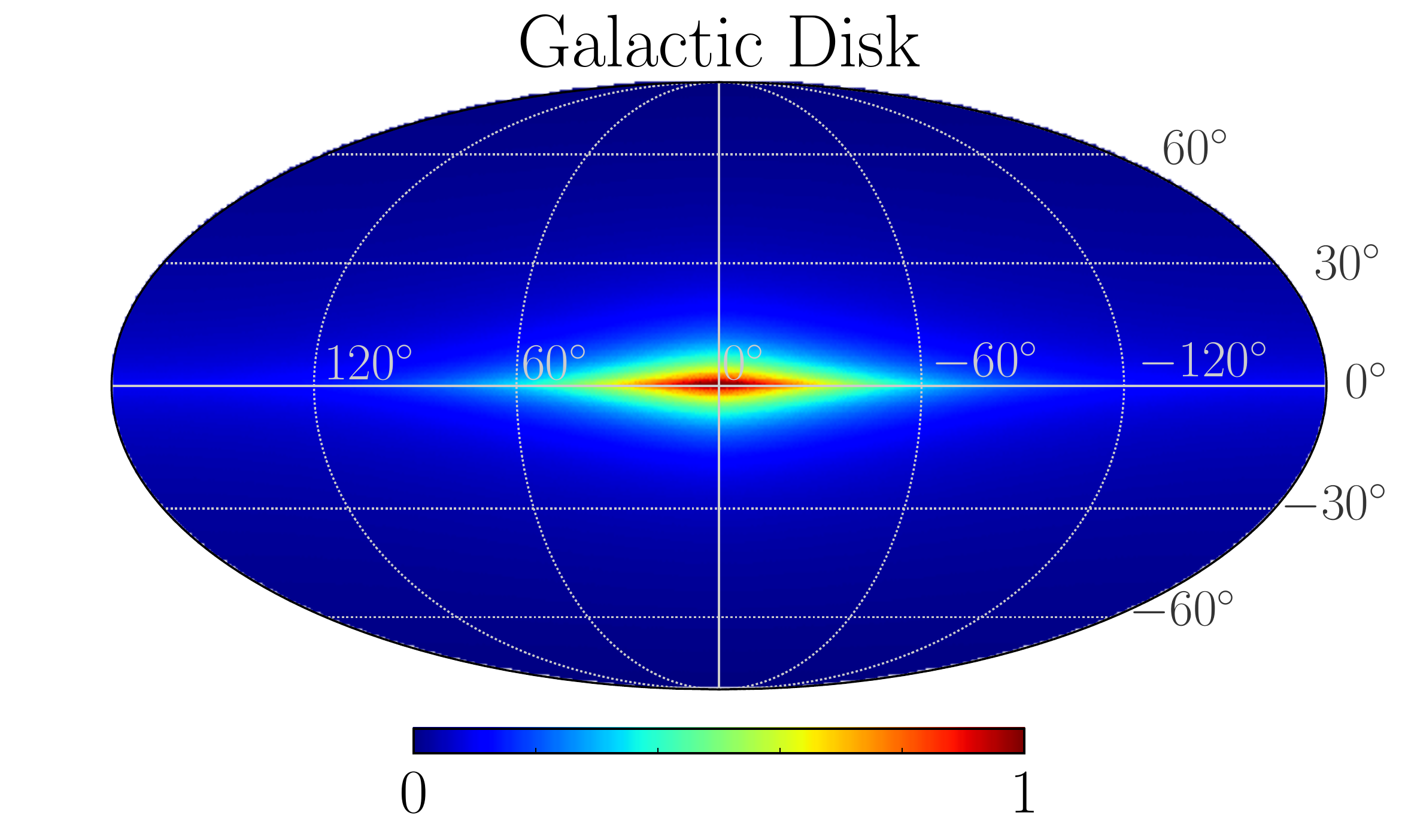}
\includegraphics[scale=0.245]{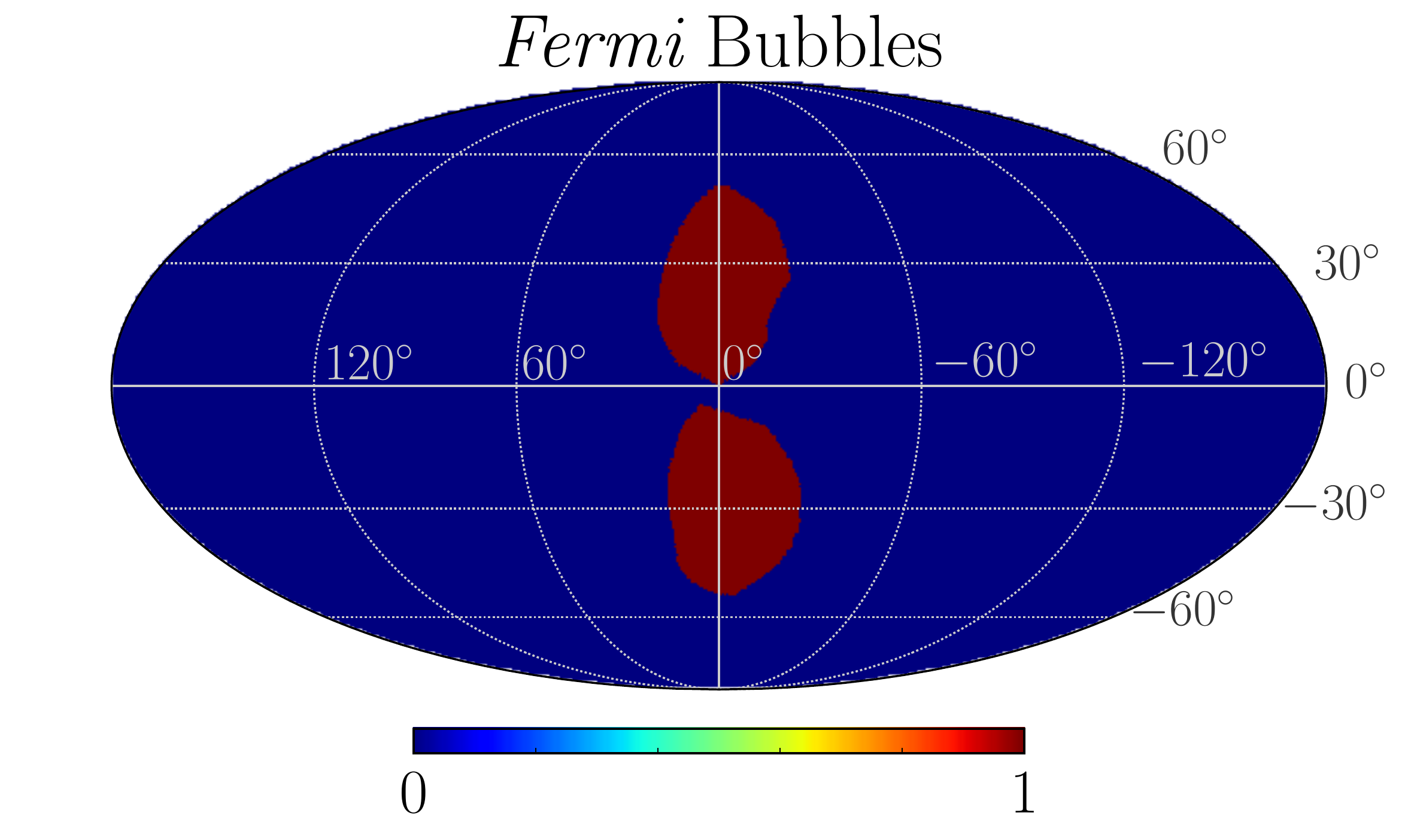}
\includegraphics[scale=0.245]{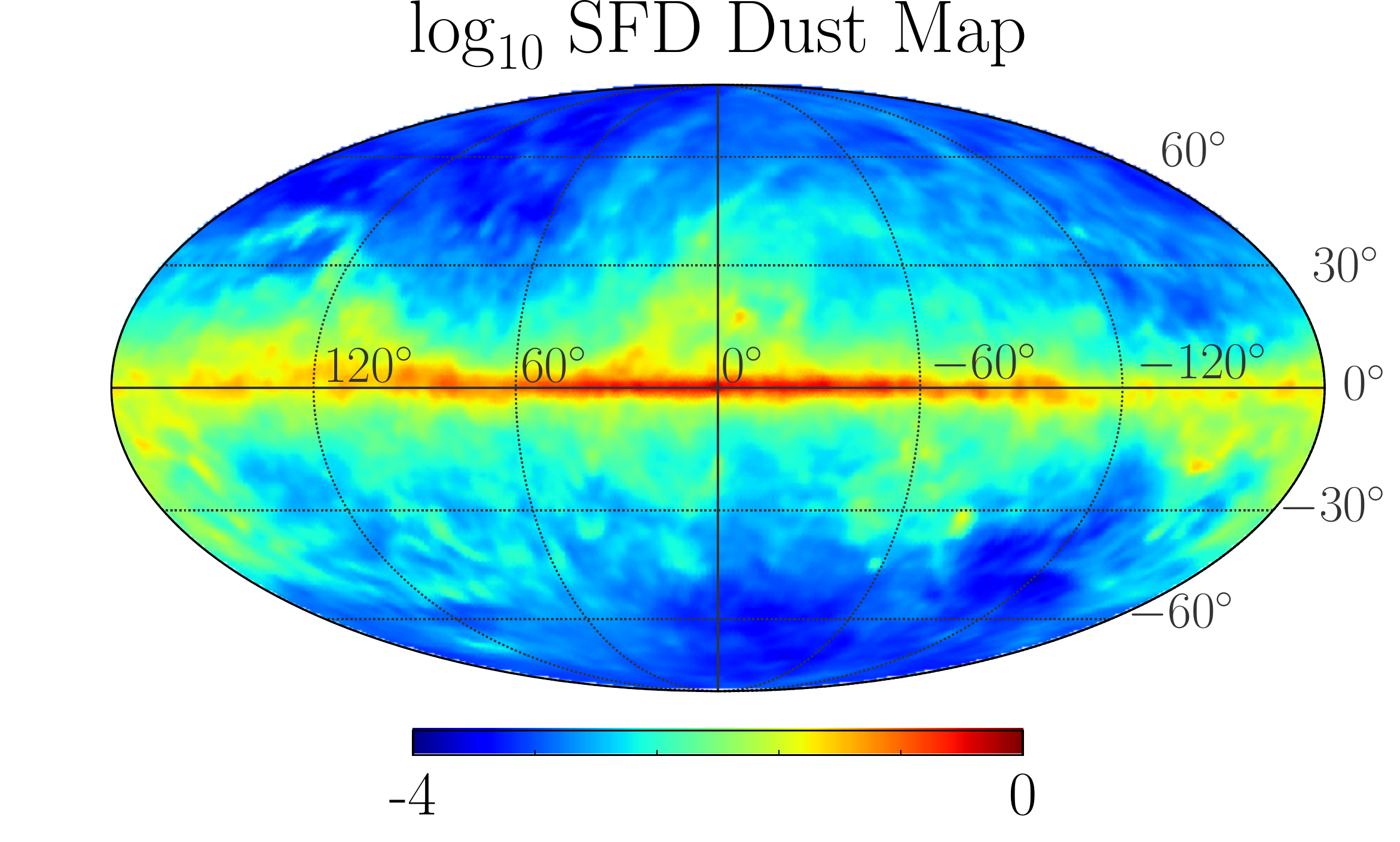}
\end{center}
\vspace{-0.5cm}
\caption{Three forms of point source distribution considered in this work, in addition to a purely isotropic distribution.
From left to right these are a model for the Galactic disk, the \textit{Fermi} bubbles~\cite{Su:2010qj}, and the SFD dust map~\cite{Schlegel:1997yv}.
In effect these are maps of $T_p^{\rm PS}$ in galactic coordinates, and the normalizations are arbitrary.
All maps are on a linear scale except for the SFD dust map, where we use a log color axis to emphasize the more detailed structure within this map.
See text for further description.
}
\label{fig:Temps}
\vspace{-0.4cm}
\end{figure*}

In detail, the Poissonian hypothesis is a model consistent of two Poisson templates: one following the dominant background contribution, given by $T_p^{\rm bkg}$, and the other accounting for the possibility that for a given spatial distribution $T_p^{\rm PS}$, the data may have a diffuse rather than unresolved point source origin.
For this second source of emission, as the flux is diffuse it is better described by Poisson statistics, and thus follows a template $\mathcal{E}_p T_p^{\rm PS}$, where the extra factor of $\mathcal{E}_p$ is required to convert to a counts map.
In this case we can write down a likelihood function as described above, and from the data we can construct the marginal likelihood as follows,
\be
\mathcal{L}_0(d) = \int \mathrm{d} {\boldsymbol \theta}\, \mathcal{L}_{\rm P}(d | {\boldsymbol \theta}) p({\boldsymbol \theta})\,,
\ee
where the subscript $0$ indicates we are using this as our null hypothesis, and the subscript P on the likelihood identifies this as the appropriate form for the Poissonian hypothesis.
In detail, $\mathcal{L}_{\rm P}(d | {\boldsymbol \theta})$ can be determined directly from Eq.~\ref{eq:Poisppk}, as we are only considering Poissonian models.
In addition we have introduced $p({\boldsymbol \theta})$, which represents the priors on the parameters.
These are given in Table~\ref{tab:priors}, where $A^{\rm bkg}$ and $A^{\rm P}$ are the normalizations of the templates $T_p^{\rm bkg}$ and $\mathcal{E}_p T_p^{\rm PS}$ respectively, and so we see this is a two parameter model.

The non-Poissonian hypothesis is derived from the null hypothesis, except that we append one further model: a non-Poissonian template following the spatial template $T_p^{\rm PS}$.
As we will take a singly-broken power law to describe the source-count distribution, this hypothesis is a six parameter model.
Once more we can form the marginal likelihood using
\be
\mathcal{L}_1(d) = \int \mathrm{d} {\boldsymbol \theta}\, \mathcal{L}_{\rm NP}(d | {\boldsymbol \theta}) p({\boldsymbol \theta})\,,
\ee
where again the priors are given in Table~\ref{tab:priors}.
Note that all priors are uniform, except for $A$ and $F_b$, which are log uniform.
Also the prior for $A^{\rm P}$ is allowed to float negative, in order to allow the fit to conserve the total amount of flux when evaluating the non-Poissonian hypothesis.
To justify this choice, recall that $T_p^{\rm bkg}$ was constructed by scrambling the real data.
Accordingly, if there is a detectable point source population within the data, the flux from these sources would be picked up by the non-Poissonian template, but also be present in $T_p^{\rm bkg}$, and therefore double counted.
A negative $A^{\rm P}$ can then be loosely thought of as subtracting that flux off, but done in such a way that the combined Poissonian template $A^{\rm bkg} T_p^{\rm bkg} + A^{\rm P} \mathcal{E}_p T_p^{\rm PS} \geq 0$ in every pixel.
We emphasize that the model used for non-Poissonian hypothesis includes both Poissonian and non-Poissonian templates, and thus the full generating function in Eq.~\ref{eq:PandNPgen} is required.

Model selection between these two hypotheses is considered through the use of the Bayes factor
\be
\mathcal{B}_{\rm NP/P}(d) = \frac{\mathcal{L}_1(d)}{\mathcal{L}_0(d)} = \frac{\int \mathrm{d} {\boldsymbol \theta}\, \mathcal{L}_{\rm NP}(d | {\boldsymbol \theta}) p({\boldsymbol \theta})}{\int \mathrm{d} {\boldsymbol \theta}\, \mathcal{L}_{\rm P}(d | {\boldsymbol \theta}) p({\boldsymbol \theta})}\,.
\label{eq:TS}
\ee
From this definition, it can be seen that the Bayes factor is a summary statistic: it integrates over all possible forms of the source-count function through the integral over parameters.
As such it provides a gross evaluation as to whether a point-source population is preferred by the data, rather than singling out any particular $\mathrm{d}N/\mathrm{d}F$.
As a summary statistic, it can also serve a secondary role as a test statistic.
Our expectation for the value of the Bayes factor can be calibrated through use of frequentist methods such as calculating the $p$-value of the Bayes factor.

To compare more specific model hypotheses, we introduce the \textit{pointwise likelihood ratio}, defined as
\be
\mathcal{M} \left(d ; {\boldsymbol \phi} \right) = \frac{\tilde{\mathcal{L}}_{\rm NP} \left(d \vert {\boldsymbol \phi} \right) \pi_{\textrm{NP}}}{\mathcal{L}_0(d)\pi_{\textrm{P}}+\mathcal{L}_1(d) \pi_{\textrm{NP}}}\,.
\label{eq:plr}
\ee
where $\pi_{\textrm{P}}$ and $\pi_{\textrm{NP}}$ are, respectively, the model priors for the Poissonian and non-Poissonian hypotheses, which have been chosen to be equal for this presentation of the results.
Here ${\boldsymbol \phi} = \left\{ N^{\rm PS}, \bar{F}^{\rm PS} \right\}$ represents the expected number of point sources across the whole sky and the expected flux per source at this location in model space, both of which were defined in~Eq.~\ref{eq:NbarF}.
In this expression, $\tilde{\mathcal{L}}_{\rm NP} \left(d \vert {\boldsymbol \phi} \right)$ is similar to $\mathcal{L}_{\rm NP} \left(d \vert {\boldsymbol \theta} \right)$, except that $n_1$ and $n_2$ have been marginalized over.
Intuitively, $\mathcal{M} \left(d ; {\boldsymbol \phi} \right)$ should be thought of as the probability for the NPTF model at this particular value of $\left\{ N^{\rm PS}, \bar{F}^{\rm PS} \right\}$, compared to the probability for an equally weighted mixture of the Poissonian and non-Poissonian models.
This definition was motivated by the need for a metric that handles both the high and low signal strength regimes.
When $\mathcal{B}_\mathrm{NP/P} \ll 1$, $\mathcal{M}$ is approximately the model posterior conditioned on $\boldsymbol{\phi}$; while in the $\mathcal{B}_\mathrm{NP/P} \gg 1$ regime, $\mathcal{M}$ is approximately the ratio between the posterior and prior: $p(\boldsymbol{\phi}|d)/p(\boldsymbol{\phi})$.

Before concluding this section, we have said a number of times we will exploit the power of the NPTF to test several forms for the spatial dependence of the point-source population, specified by $T_p^{\rm PS}$.
In addition to isotropically distributed sources, where $T_p^{\rm PS} \propto 1$, we will consider three additional templates that consider the possibility of point sources distributed within the Milky Way, all of which are shown in Fig.~\ref{fig:Temps}.
The galactic disk template is used as a generic model for sources distributed according to the disk of the Milky Way.
This is the line-of-sight integrated emission from a disk with a source density that scales exponentially in radius and distance from the plane, with a scale height of 0.3 kpc and a scale radius of 5 kpc.
Next we consider sources distributed following the \textit{Fermi} bubbles~\cite{Su:2010qj}, large structures observed in gamma-rays extending perpendicularly from the galactic disk.
Although the emission observed from the bubbles so far appears diffuse, neutrino emission from the recently discovered small scale gas clouds within the bubbles~\cite{2018ApJ...855...33D}, could lead to point like sources.
Finally we consider the Schlegel, Finkbeiner, and Davis (SFD) dust map~\cite{Schlegel:1997yv}, which provides a 2 dimensional distribution of dust within the Milky Way, mapped out using the reddening of starlight.
This distribution is interesting to consider as dust and gas tend to have similar spatial distributions, so the map is a proxy for the distribution of hydrogen in the Milky Way.
The hydrogen is a target for cosmic-ray proton to collide with and form pions.
The neutral pions then decay to photons, and indeed the SFD dust map can be seen clearly in the \textit{Fermi} gamma-ray data.
If these same interactions produce higher energy pions, then the charged variants could produce neutrinos at IceCube, making this an interesting spatial distribution to consider.
As the target dust has a diffuse distribution throughout the Milky Way, in order for this template to describe a point-source population, the sources of the cosmic-ray protons would need to be point like, as this would then imprint a point-source-like distribution into the neutrino data.
Note that we have chosen not to divide our templates between the northern and southern sky, even though this is commonly done for extragalactic point source searches, see for example \textcite{Aartsen:2016oji}.
Usually, this distinction between the northern and southern sky is imposed due to the different backgrounds that dominate in each hemisphere; in the northern sky, the main background is atmospheric neutrinos, whereas in the southern sky instead atmospheric muons dominate.
Nonetheless, for the isotropic case we show both the sensitivity and results for the northern sky in App.~\ref{sec:NorthIso}.
There we will see that the reach for the restricted case is only slightly enhanced, justifying our choice to focus on the full sky.

%%%%%%%%%%%%%%%%%%%%%%%%%%%%%%%
\section{Expected Sensitvity}\label{sec:ExpSens}
%%%%%%%%%%%%%%%%%%%%%%%%%%%%%%%

\begin{figure}[t]
\begin{center}
\includegraphics[scale=0.35]{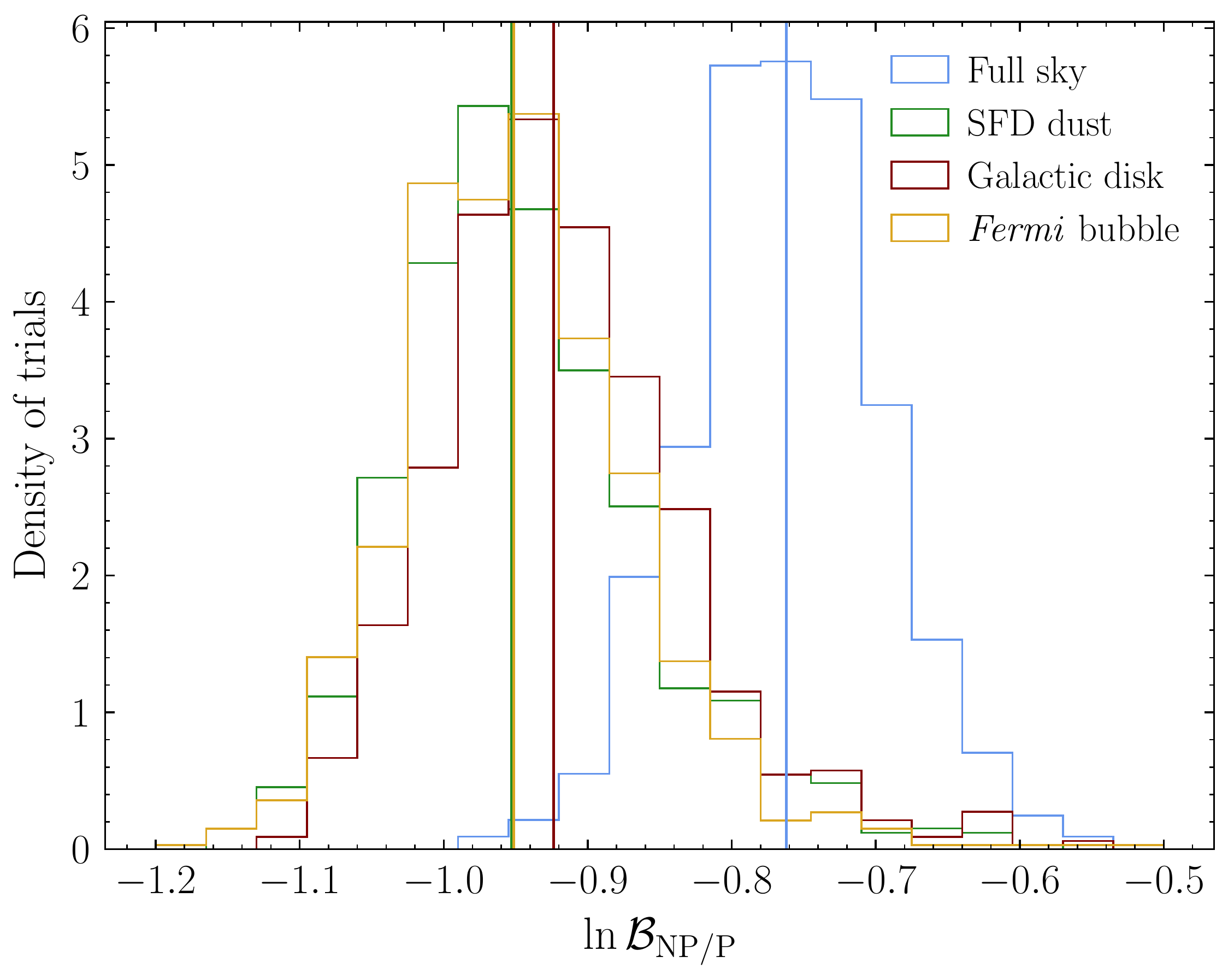}
\end{center}
\vspace{-0.5cm}
\caption{The distribution of the test statistic, as given in~Eq.~\ref{eq:TS}, under the background only hypothesis for four different signal states.
These distributions, formed from 1000 trials, are used for establishing sensitivities and $p$-values.
In particular, the vertical lines represent a $p$-value of 0.5.
Here, each trial is formed by scrambling the data in right ascension.
See text for details.
}
\label{fig:NullDist}
\vspace{-0.4cm}
\end{figure}

\begin{figure*}[t]
\begin{center}
\includegraphics[scale=0.35]{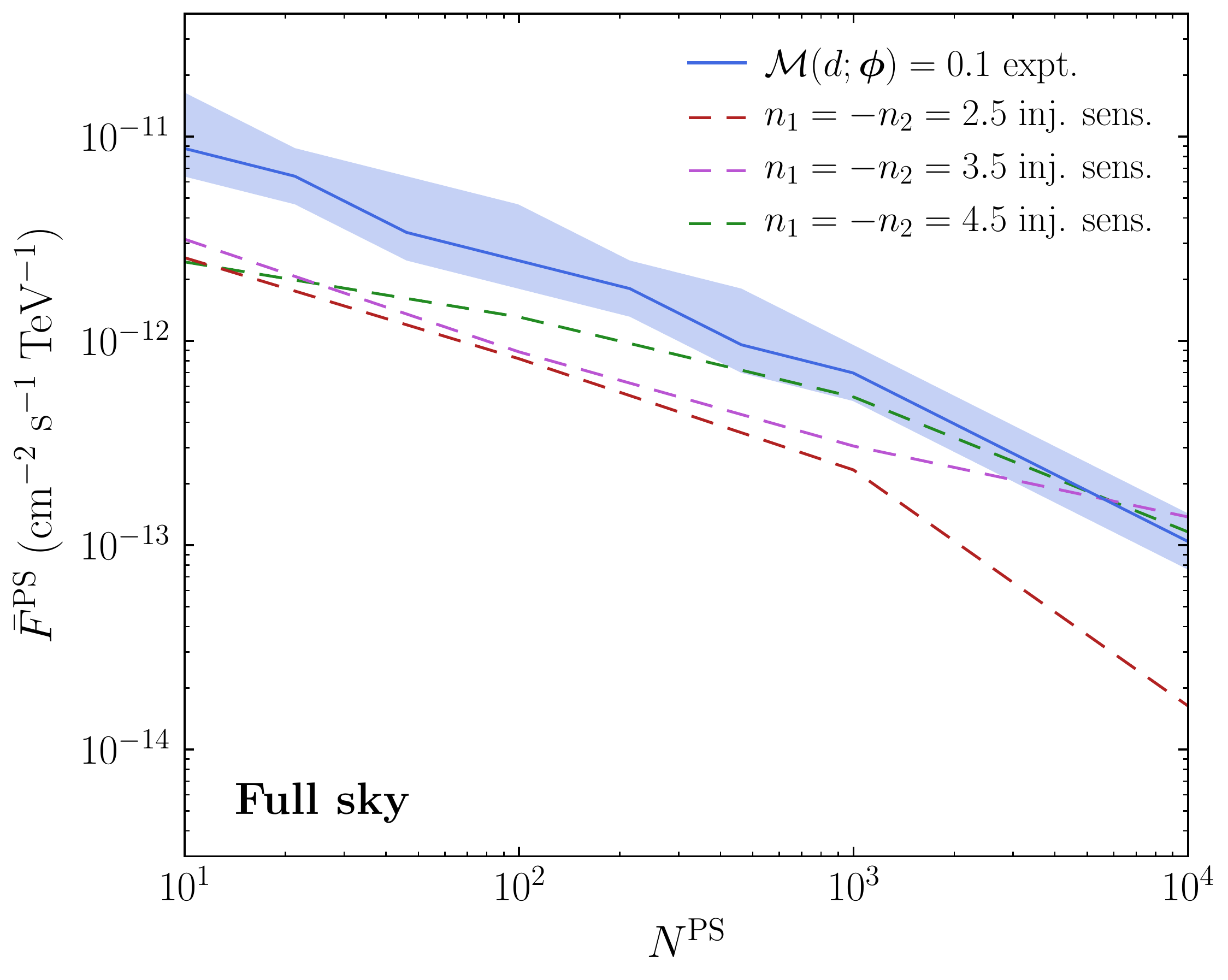} \hspace{0.5cm}
\includegraphics[scale=0.35]{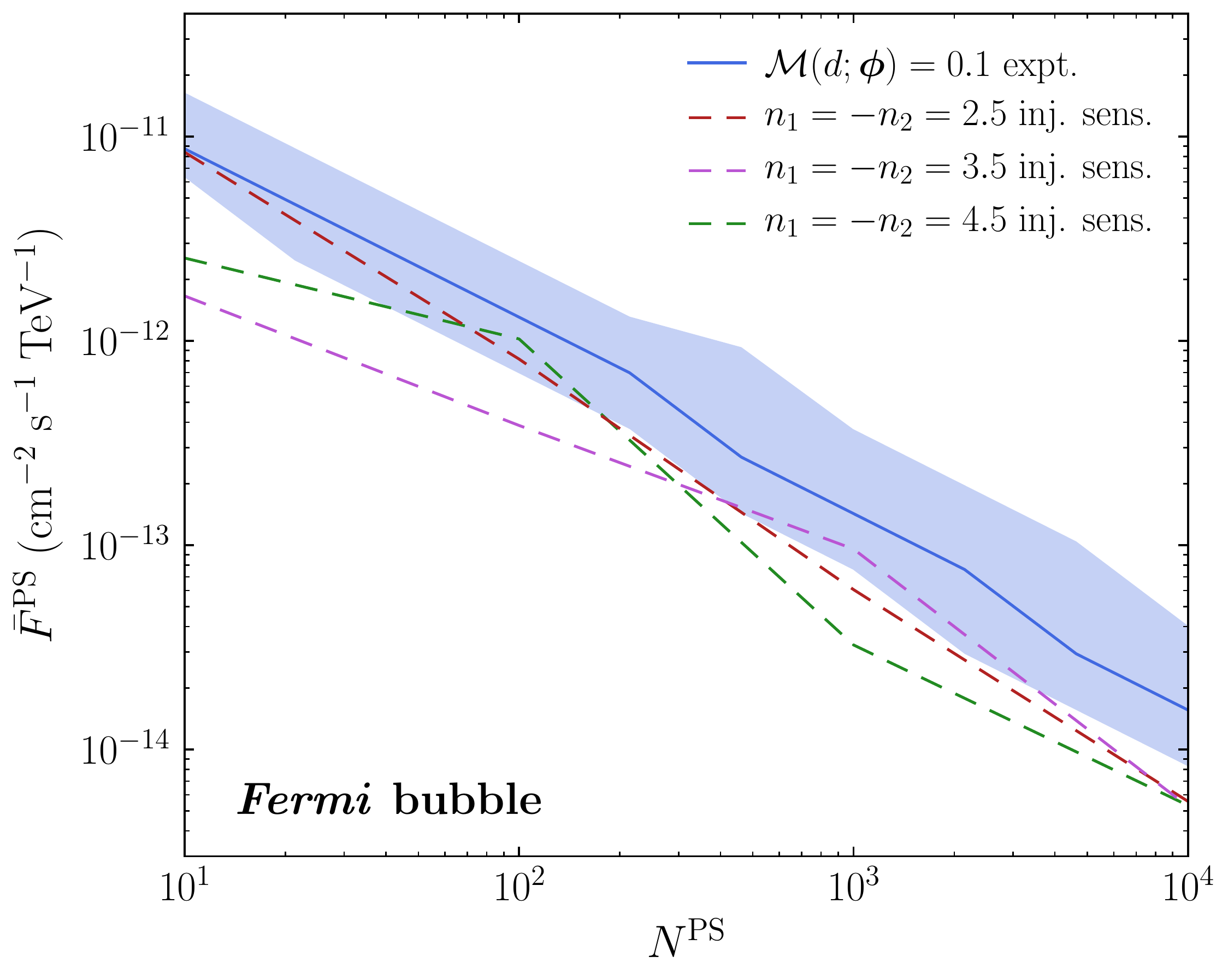} \\ \vspace{0.5cm}
\includegraphics[scale=0.35]{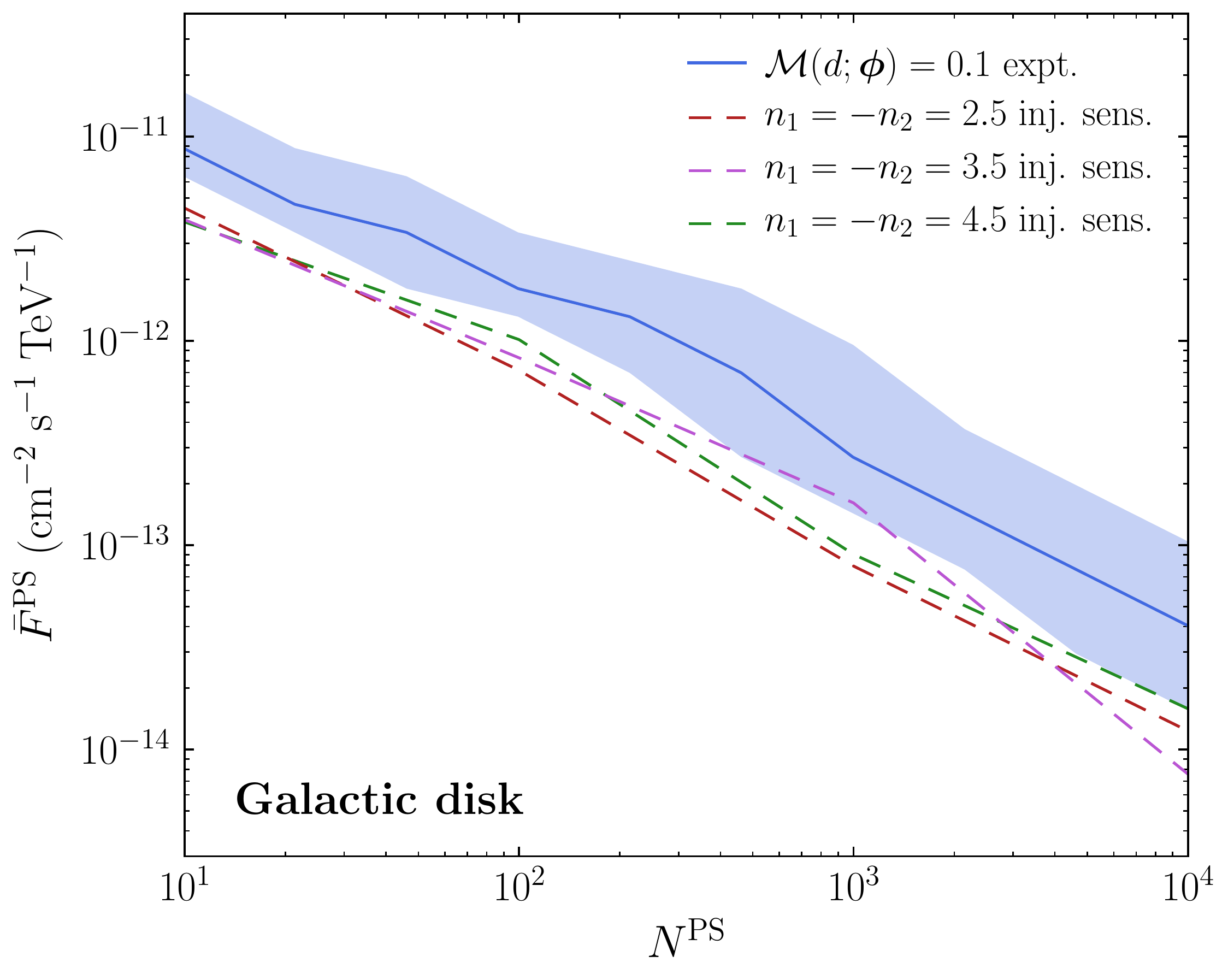} \hspace{0.5cm}
\includegraphics[scale=0.35]{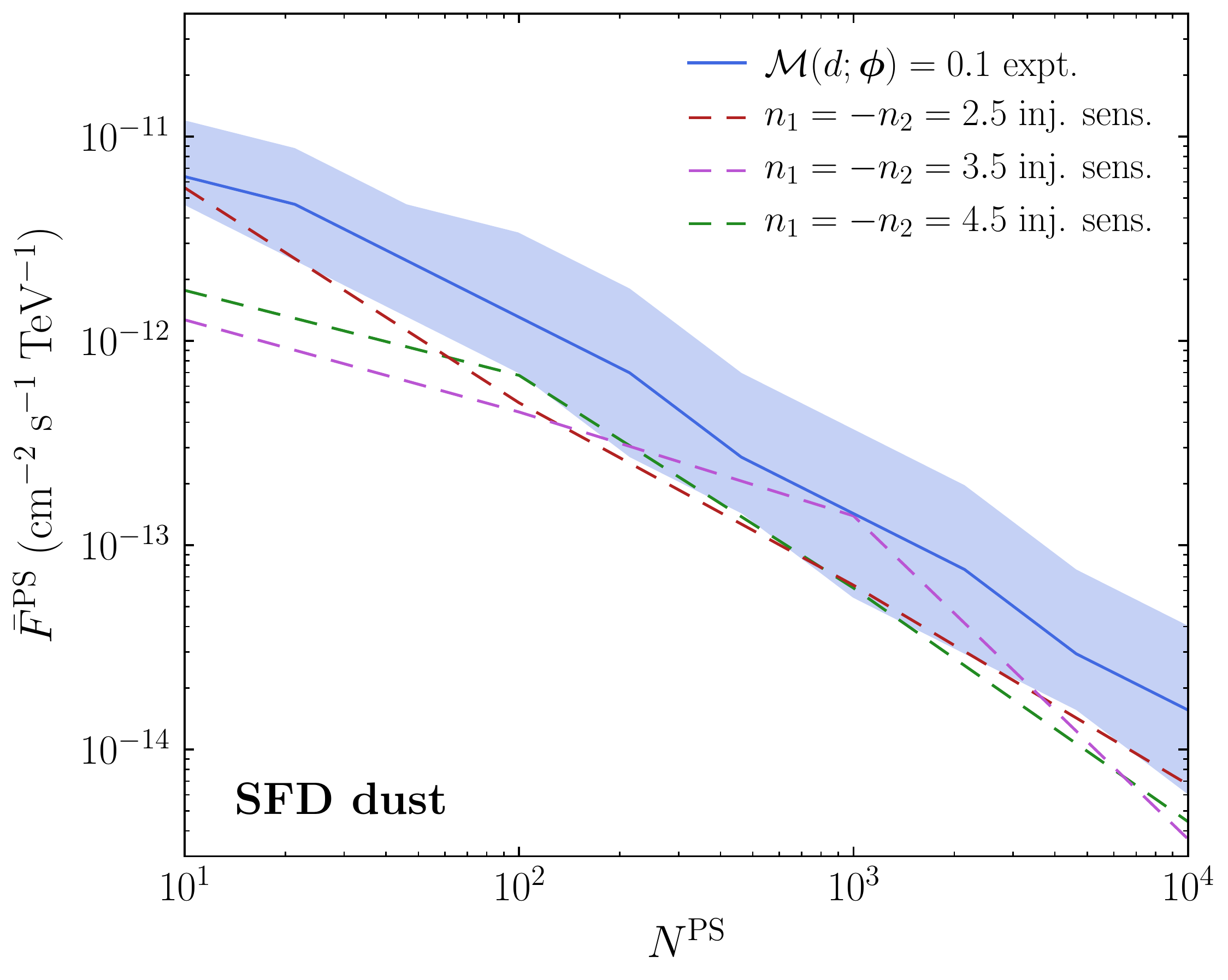}
\end{center}
\vspace{-0.5cm}
\caption{The expected sensitivity and limit as a function for the four different spatial templates considered for the point source distributions: isotropic extragalactic sources over the full sky (top left), \textit{Fermi} bubbles (top right), SFD dust (bottom right), and galactic disk (bottom left).
In each case, the sensitivity is shown as the dashed curves for three different shapes of the source-count function.
The median expected limit derived using $\mathcal{M}(d;{\boldsymbol \phi}) = 0.1$ is shown in blue, as well as the associated 10th and 90th percentiles from the distribution.
See text for details.
}
\label{fig:sensitivity}
\vspace{-0.4cm}
\end{figure*}

\begin{figure*}[t]
\begin{center}
\includegraphics[scale=0.35]{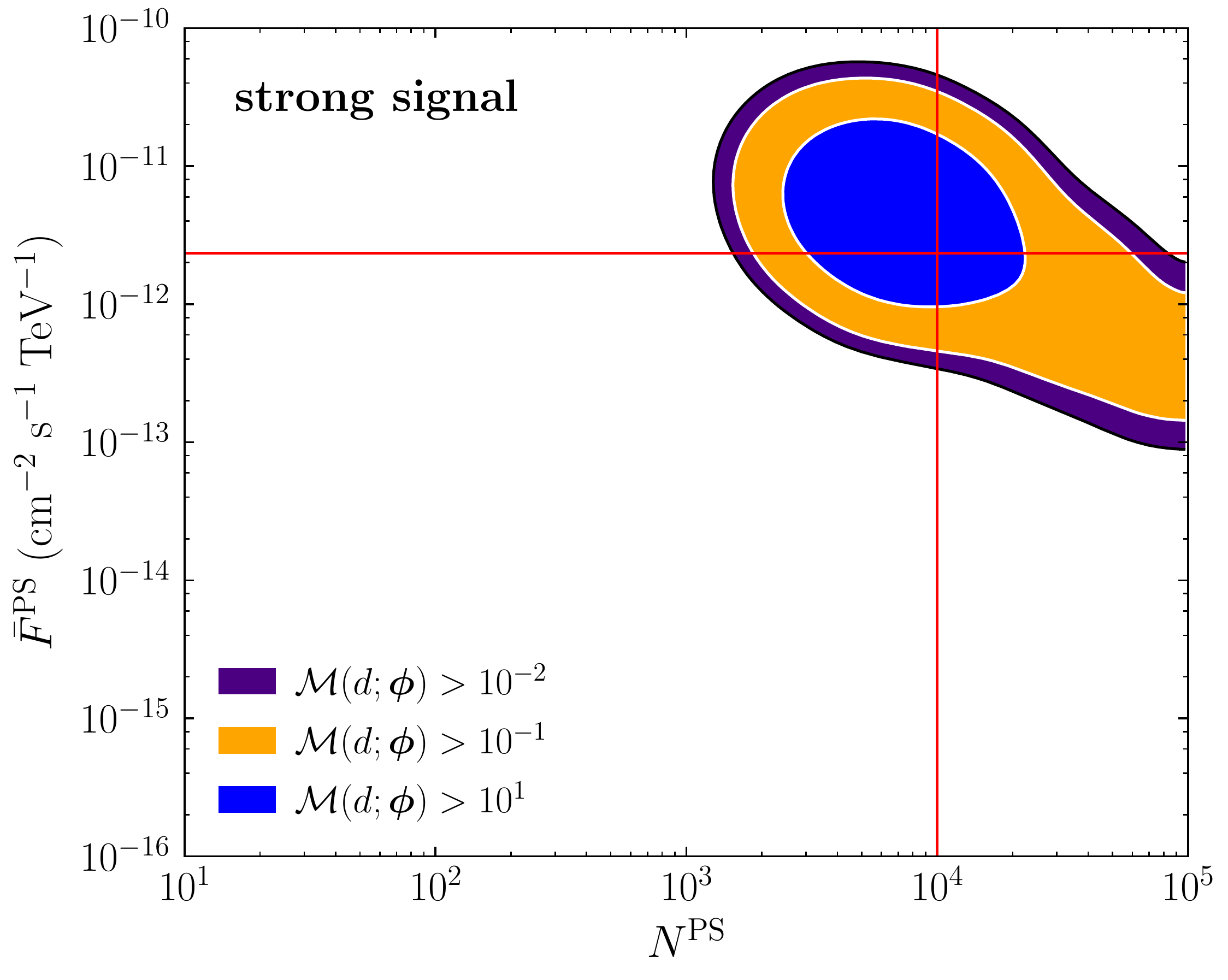} \hspace{0.1cm}
\includegraphics[scale=0.35]{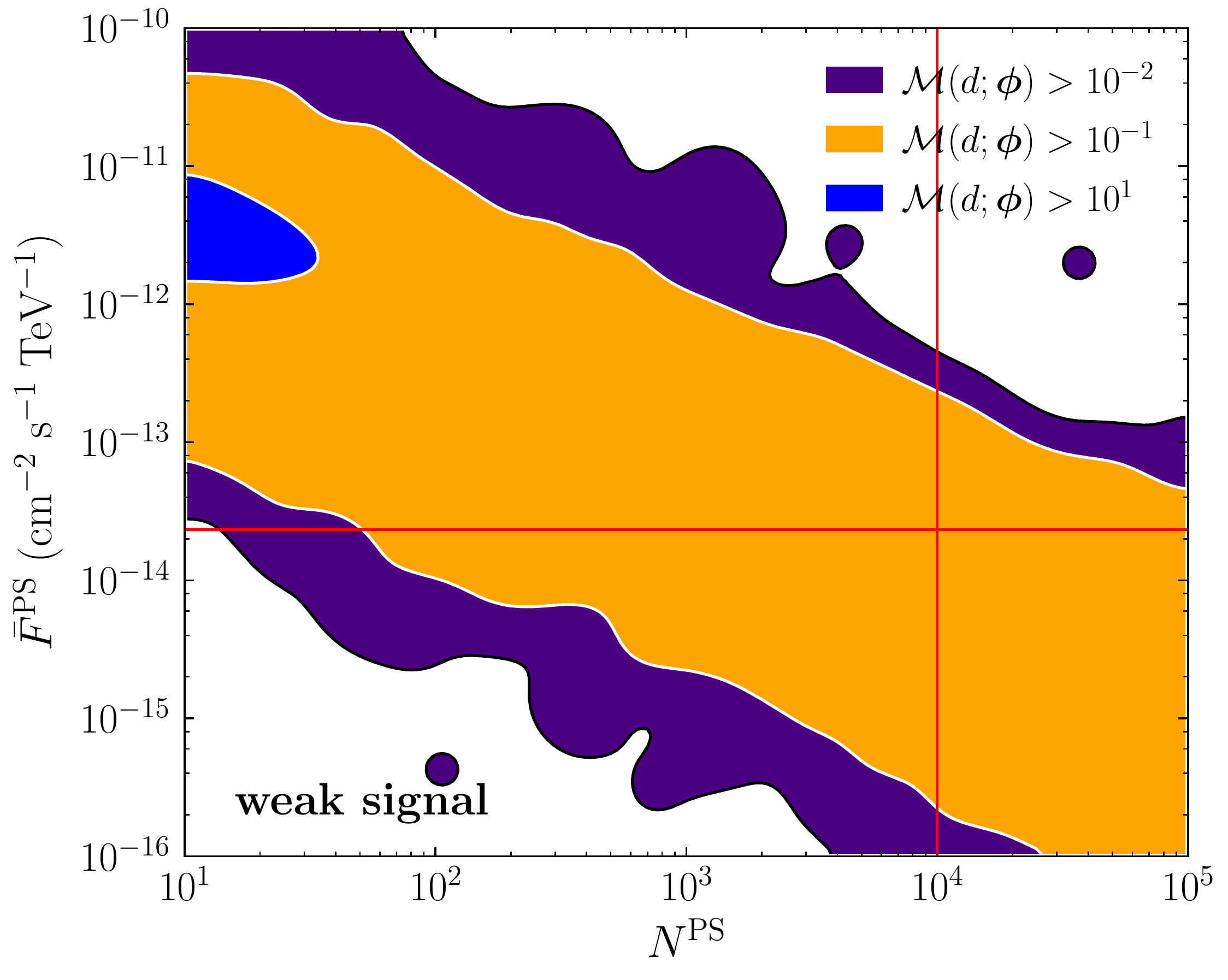} \hspace{0.1cm}
\includegraphics[scale=0.35]{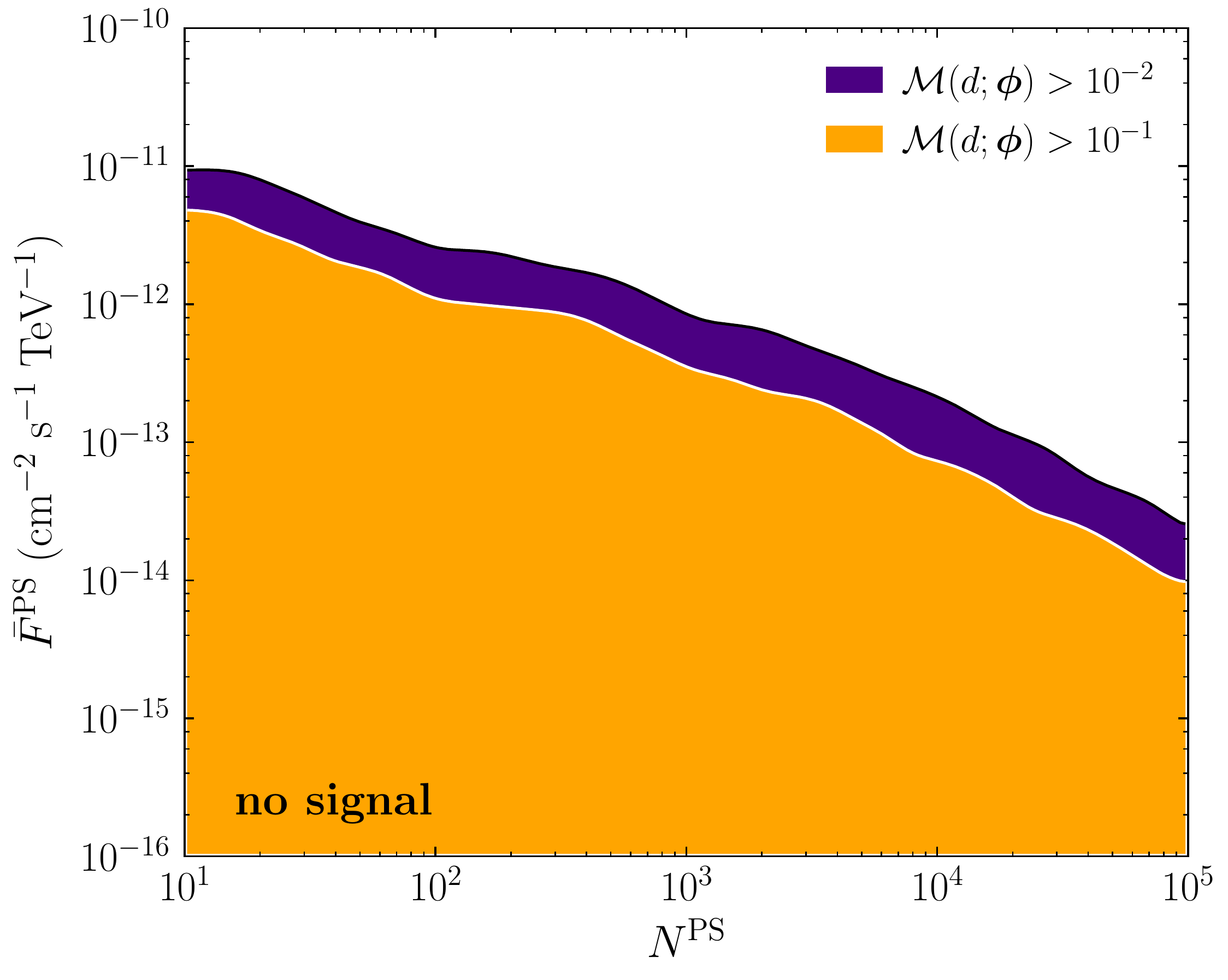}
\end{center}
\vspace{-0.5cm}
\caption{The map of $\mathcal{M}(d;{\boldsymbol \phi})$ for three different datasets including injected point-source populations.
In each of the cases, the parameters of the injected population are shown with the red lines.
Only in the case of a strong signal is this method able to identify the specific location in this reduced parameter space of the signal.
See text for details.
}
\label{fig:injsig}
\vspace{-0.4cm}
\end{figure*}

\begin{figure}[t]
\begin{center}
\includegraphics[scale=0.35]{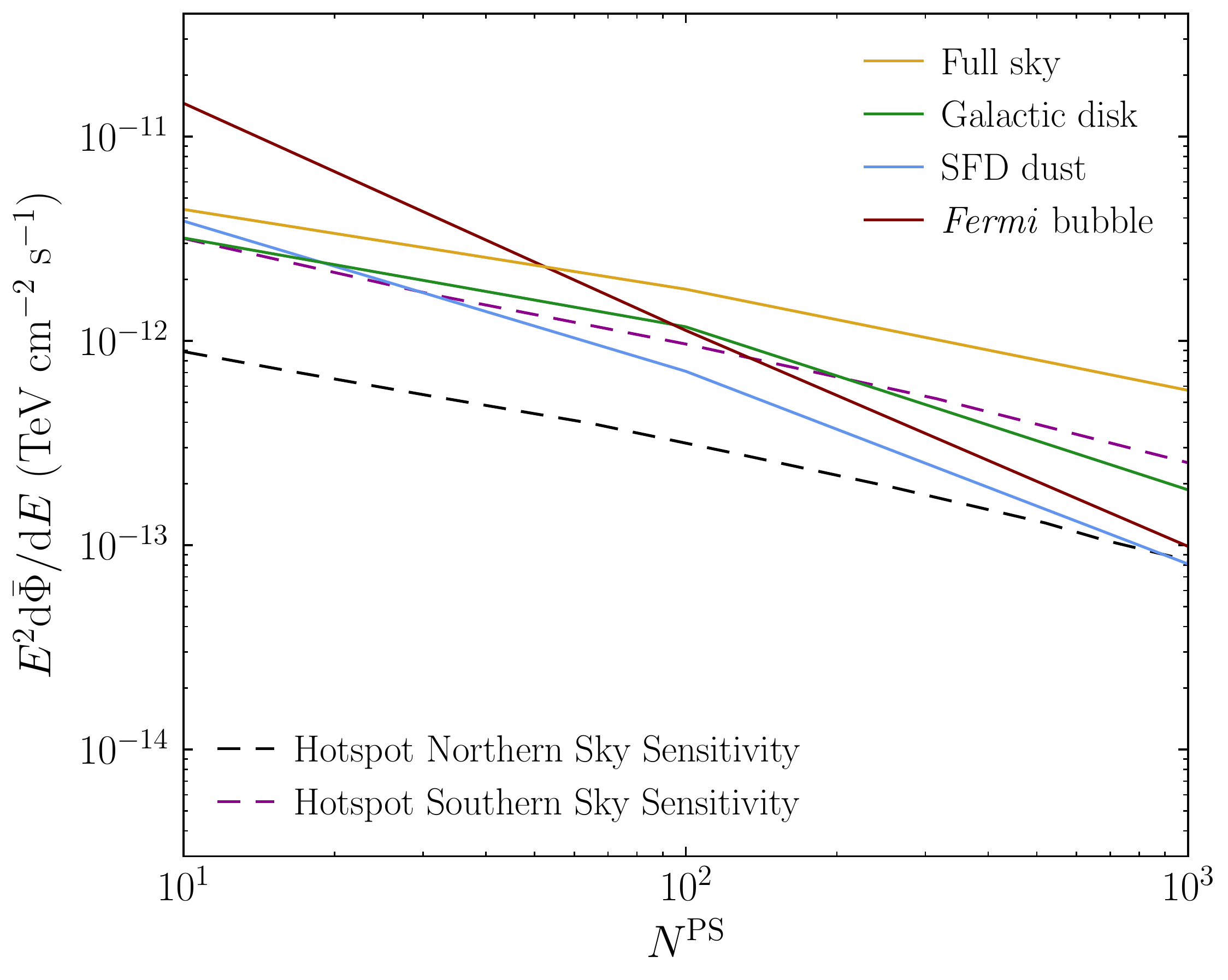}
\end{center}
\vspace{-0.5cm}
\caption{The expected sensitivities for four templates when $n_1 = -n_2 = 20$ is fixed, creating an approximation of an equal-flux population of sources. 
Northern and southern sky hotspot sensitivities for populations of equal-flux source from \textcite{Aartsen:2016oji} are shown along-side.
The hotspot analysis represents a traditional approach to point source detection, where a ``hotspot" is defined as a source that is individually detected at 3$\sigma$ global significance.
See text for details.
}
\label{fig:1609_comparison}
\vspace{-0.4cm}
\end{figure}

Using the techniques and statistical framework described in the previous section, we now turn to estimating the expected reach of this technique using Monte Carlo simulations.
We will consider both the case of setting limits and quantifying thresholds for a discovery of a point-source population.
As we use techniques generally motivated by Bayesian statistics, part of the aim of this section is to help develop intuition for what discovery and limit setting looks like in our framework.
Nevertheless, the primary output of a Bayesian analysis is the posterior, and as mentioned we make this publicly available,\footnote{\url{https://icecube.wisc.edu/science/data/NPTF_7yr_posterior}} describing the details in App.~\ref{sec:PubPost}.

To begin with, we consider the expected limit sensitivity for the analysis.
The sensitivity is determined by comparing how our test statistic, the Bayes factor given in~Eq.~\ref{eq:TS}, is distributed over many trials for each of the signal plus background and background only hypothesis.
From these distributions, our sensitivity to a given model is defined as when 90\% of the signal distribution is above 50\% of the background case.
For example, the background only distribution of the natural log of our test statistic, generated from 1000 trials, is shown in Fig.~\ref{fig:NullDist} for each of our four signal templates.
The trials are generated by taking the real data, but scrambling the right ascension of each event, with a different scrambling for each trial.
This background only distribution can also be used to establish $p$-values, and indeed the definition of sensitivity is equivalent to requiring the $p$-value for 90\% of the signal distribution to be less than 0.5.\footnote{A similar procedure can be used to establish the expected discovery sensitivity, which is defined as when 50\% of the signal distribution has a $p$-value less than $2.87 \times 10^{-7}$, the threshold usually referred to as a 5$\sigma$ discovery.
Determining the associated test statistic that corresponds to such a small $p$-value requires generating a large number of background only trials, and as we find no significant evidence for a signal in the present analysis we have not quantified the discovery threshold.
}
Note that as sensitivity is defined in terms of the distributions, there is no statistical variation in its value, as opposed to say a frequentist 90\% confidence limit.
Where the sensitivity threshold occurs will not be a unique point in the signal parameter space, established by $\{ N^{\rm PS}, \bar{F}^{\rm PS}, n_1, n_2 \}$.
If, however, we fix three of the parameters, for example $N^{\rm PS}$, $n_1$, and $n_2$, then we can define the sensitivity as a function of $\bar{F}^{\rm PS}$.
With this in mind, in Fig.~\ref{fig:sensitivity} we show the sensitivity to $\bar{F}^{\rm PS}$ as a function of $N^{\rm PS}$, for three different values of $n_1 = - n_2$, and for the four signal templates.

In addition to the sensitivity, we define a contour where $\mathcal{M}(d; {\boldsymbol \phi})$, given in Eq.~\ref{eq:plr}, equals 0.1.
This contour is the dividing line, above which, the odds for a particular point in parameter space is no better than 1 in 10.
As $\mathcal{M}(d; {\boldsymbol \phi}) = 0.1$ varies between different datasets, we show the median, 10th, and 90th percentiles on the distribution also in Fig.~\ref{fig:sensitivity}.
Recall that $\mathcal{M}(d; {\boldsymbol \phi})$ is defined by marginalizing over $n_1$ and $n_2$.
As explained around Eq.~\ref{eq:NbarF}, we can describe an NPTF template in terms of $N^{\rm PS}$, $\bar{F}^{\rm PS}$, $n_1$, and $n_2$.
With $N^{\rm PS}$ fixed, and the indices marginalized over, the only remaining degree of freedom is the average flux per source, $\bar{F}^{\rm PS}$, which for a fixed number of sources is equivalent to the total flux associated with the non-Poissonian template.
Accordingly the limit set using this procedure effectively reduces to the weaker constraint obtained by ensuring such a population does not overproduce the observed neutrino flux, rather than drawing on the full power of the NPTF likelihood.
This explains why for a large number of sources, the expected limit is weaker for the full sky as compared with the other cases.
For the spatially restricted templates, such as the {\it Fermi} bubbles, the point sources and the majority of their associated neutrinos are forced within a smaller region on the sky compared with the full sky case, leading to a stronger limit.
As discussed earlier, all templates are applied to the full dataset, although for the isotropic case, we show the sensitivity and results restricting to the northern hemisphere in App.~\ref{sec:NorthIso}.

In addition to being useful for setting limits, we can use $\mathcal{M}(d; {\boldsymbol \phi})$ to map out the signal parameter space to determine if there are regions that are particularly preferred by the data.
In order to calibrate our analysis for this case, we will consider three different scenarios where we look for evidence of an actual point source signal.
In each of these scenarios we consider a background dataset with an additional injected point-source population.
The population is distributed isotropically over the full sky, where in all scenarios the expected number of sources is $10^4$, but each have a source-count function following a singly broken power-law, with $n_1=-n_2=2.5$.
The cases are distinguished by the value of the flux break of the source-count function, $F_b$; we consider a strong signal with $F_b = 10^{-12}$ neutrinos/cm$^2$/s/TeV, a weak signal of $F_b = 10^{-14}$ neutrinos/cm$^2$/s/TeV, and finally a case almost equivalent to no signal with $F_b = 10^{-15}$ neutrinos/cm$^2$/s/TeV.
Note the strong signal here corresponds to value that is a factor of $\sim$100 brighter than the observed diffuse flux, and the sole purpose of such a large value is to validate that the framework has been calibrated correctly.
The distribution of $\mathcal{M}(d; {\boldsymbol \phi})$ for each of these cases is shown in Fig.~\ref{fig:injsig}, and we see that only in the strong signal case is the actual injected point clearly singled out.
Nevertheless in the weak signal case, it is clear that a non-zero point-source population is preferred, but the exact location is not correctly identified.
The fit is not able to distinguish between a few bright or many dim sources.
Finally, in the no signal case the dataset is clearly consistent with no point-source population, as the injected population falls below the sensitivity of our analysis.
For reference the $\ln \mathcal{B}_{\rm NP/P}$ of the strong, weak and no signal cases, is approximately $2.3 \times 10^4$, $10.7$, and $-0.84$ respectively.

It is worth emphasizing that all results shown above, and indeed those we derive on the actual data in the next section, represent slices through the full model space.
This is the cost of reducing a four dimensional parameter space into a two dimensional limit plot.
For a specific model prediction, the full posterior is the more relevant resource.

Finally, we emphasize that a direct comparison of this method to limits set by previous IceCube analyses -- such as those in \textcite{Aartsen:2016oji} -- can not be made straightforwardly, as those limits are calculated by assuming a population of equal-flux point sources.
This kind of population can be emulated by requiring $n_1$ and $n_2$ to be fixed to a large absolute value, thus creating an approximate Dirac delta distribution differential source-count function.
In this limited region of parameter space a direct comparison is possible, and sensitivities for the four templates under consideration, using $n_1 = -n_2 = 20$, are shown in Fig.~\ref{fig:1609_comparison} along side the northern and southern sky sensitivities from \textcite{Aartsen:2016oji}.\footnote{The northern sky sensitivity from \textcite{Aartsen:2016oji} has been recalculated to account for an incorrect treatment of signal acceptance in the original publication.}

%%%%%%%%%%%%%%%%%%%%%%%%%%%%%%%
\section{Results}\label{sec:Results}
%%%%%%%%%%%%%%%%%%%%%%%%%%%%%%%

\begin{figure*}[t]
\begin{center}
\includegraphics[scale=0.35]{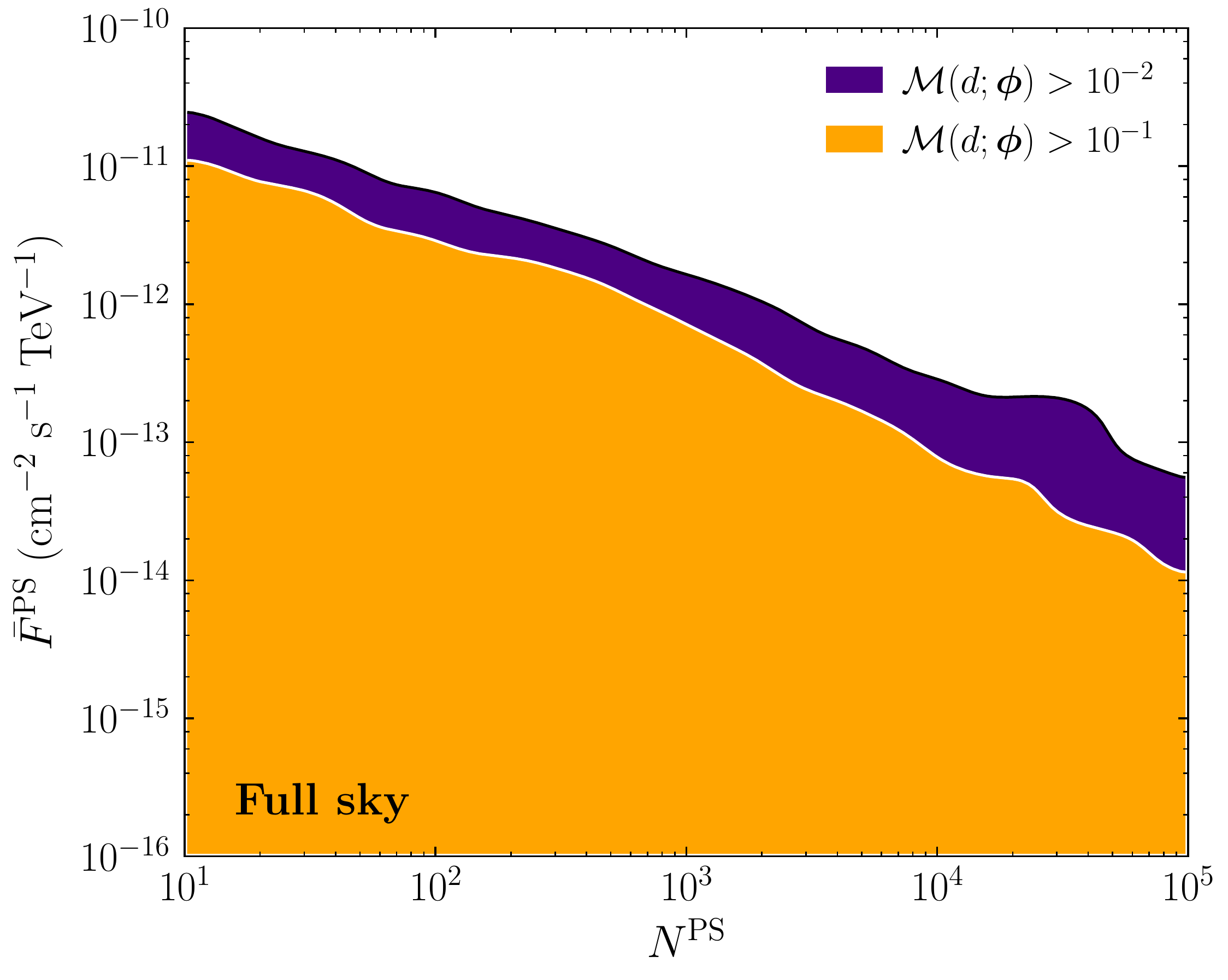} \hspace{0.5cm}
\includegraphics[scale=0.35]{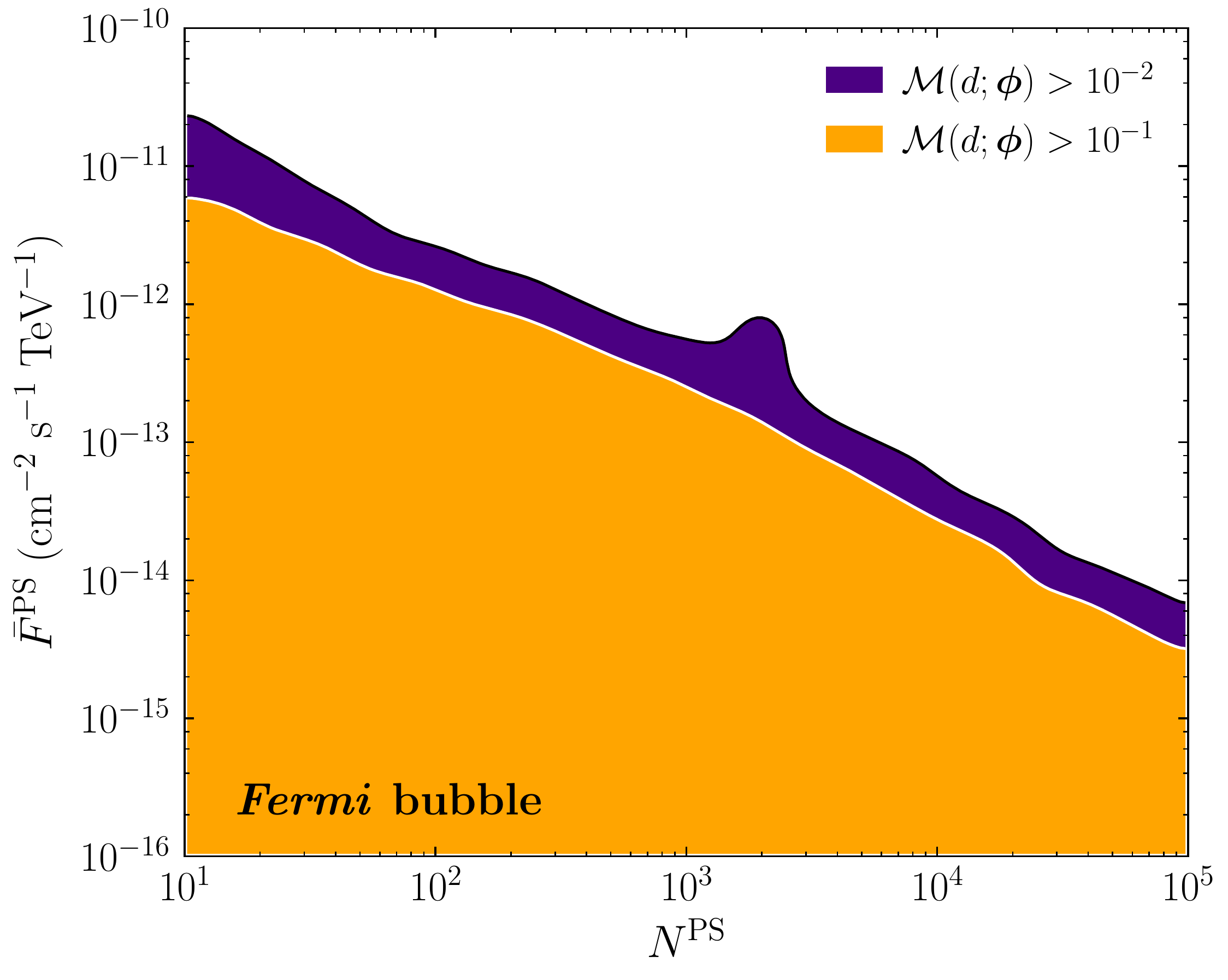} \\ \vspace{0.5cm}
\includegraphics[scale=0.35]{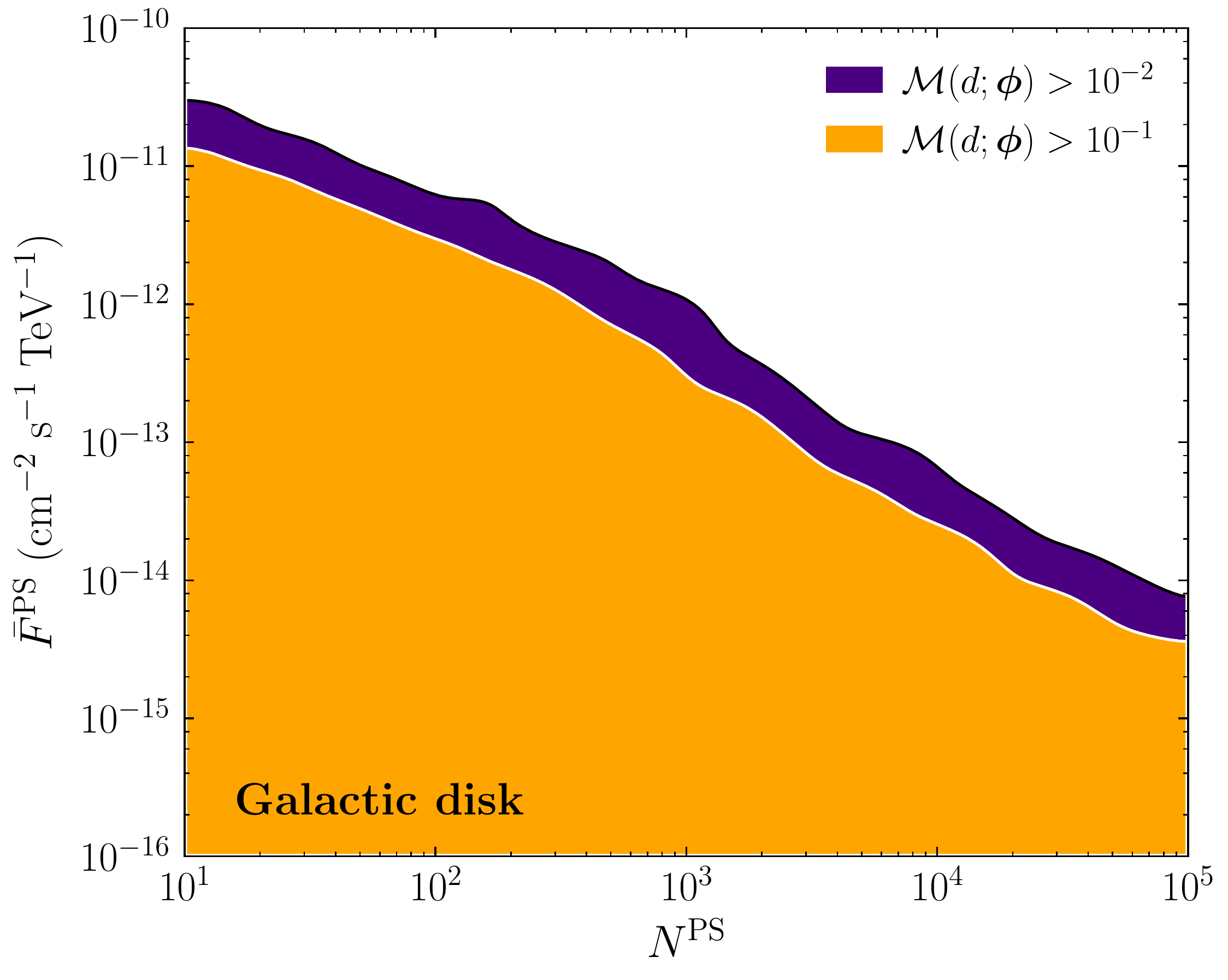} \hspace{0.5cm}
\includegraphics[scale=0.35]{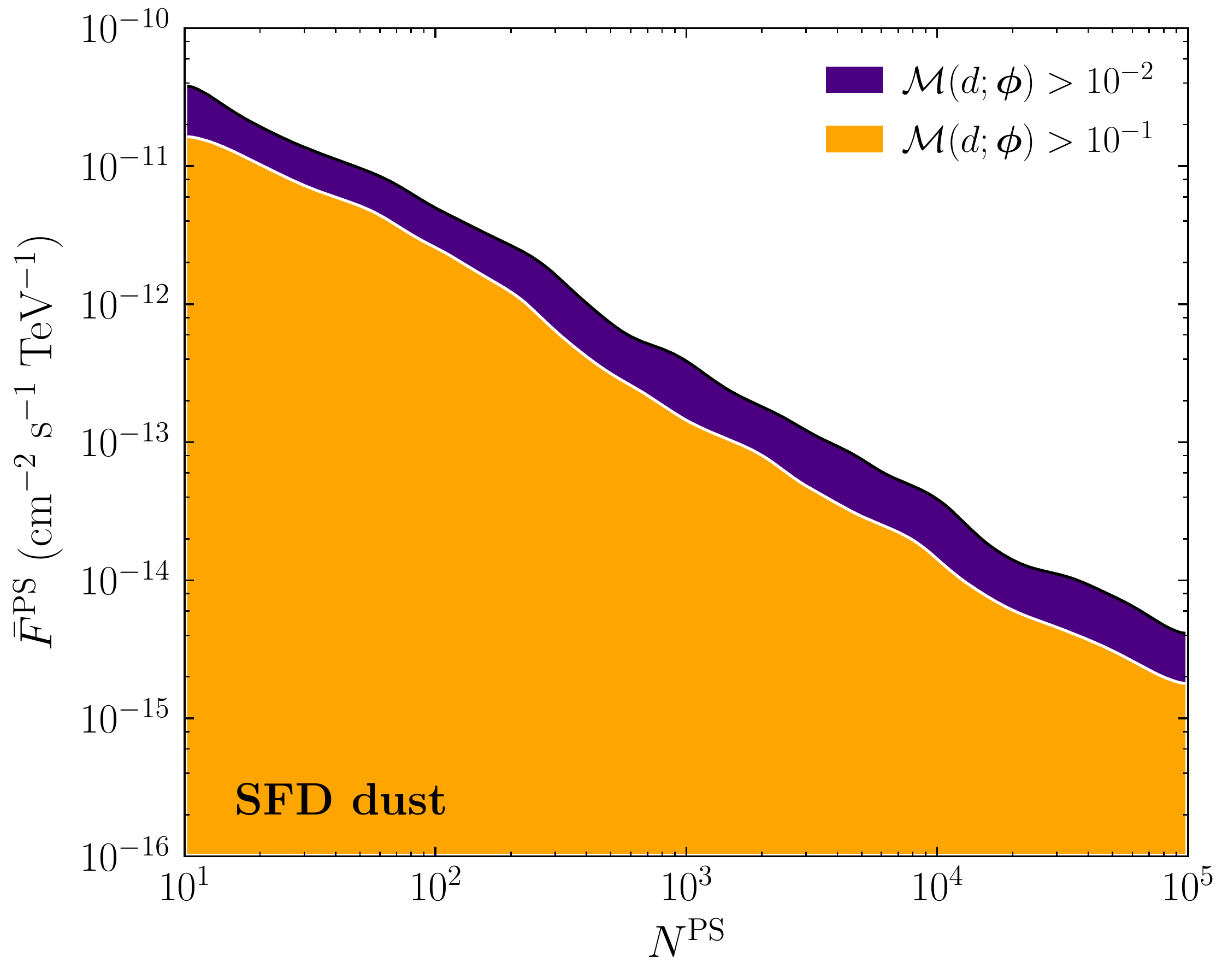}
\end{center}
\vspace{-0.5cm}
\caption{The pointwise likelihood ratio, $\mathcal{M}(d; {\boldsymbol \phi})$, for the four different point source spatial distributions considered in this work: isotropic sources over the full sky (top left), \textit{Fermi} bubbles (top right), SFD dust (bottom right), and galactic disk (bottom left).
In each case the results are consistent with the background or Poissonian hypothesis, with the most significant $p$-value of 0.33 occurring for the SFD dust map.
}
\label{fig:results}
\vspace{-0.4cm}
\end{figure*}

In this section we apply the techniques used to estimate the sensitivity discussed in the last section to the actual data.
In particular, we plot the distribution of values $\mathcal{M}(d; {\boldsymbol \phi})$ for each of our signal templates.
The results of this are shown in Fig.~\ref{fig:results}.
These plots indicate that for each of the investigated case there is no indication of a point-source population present in the data, and accordingly the results are consistent with the expected limits shown in Fig.~\ref{fig:sensitivity}.
In each case the significance can be quantified as follows:
\begin{itemize}
\item \textbf{Full sky:} $\ln \mathcal{B}_{\rm NP/P} = -0.79$, $p$-${\rm value} = 0.66$;
\item \textbf{\textit{Fermi} Bubbles:} $\ln \mathcal{B}_{\rm NP/P} = -0.94$, $p$-${\rm value} = 0.45$;
\item \textbf{SFD Dust:} $\ln \mathcal{B}_{\rm NP/P} = -0.92$, $p$-${\rm value} = 0.33$; and
\item \textbf{Galactic Disk:} $\ln \mathcal{B}_{\rm NP/P} = -0.97$, $p$-${\rm value} = 0.74$.
\end{itemize}
The $p$-values quoted here were determined from the distribution of the background only hypotheses, which were shown in Fig.~\ref{fig:NullDist}.
From the $p$-values each signal template is consistent with the Poissonian hypothesis.

\begin{figure*}[t]
\begin{center}
\includegraphics[scale=0.45]{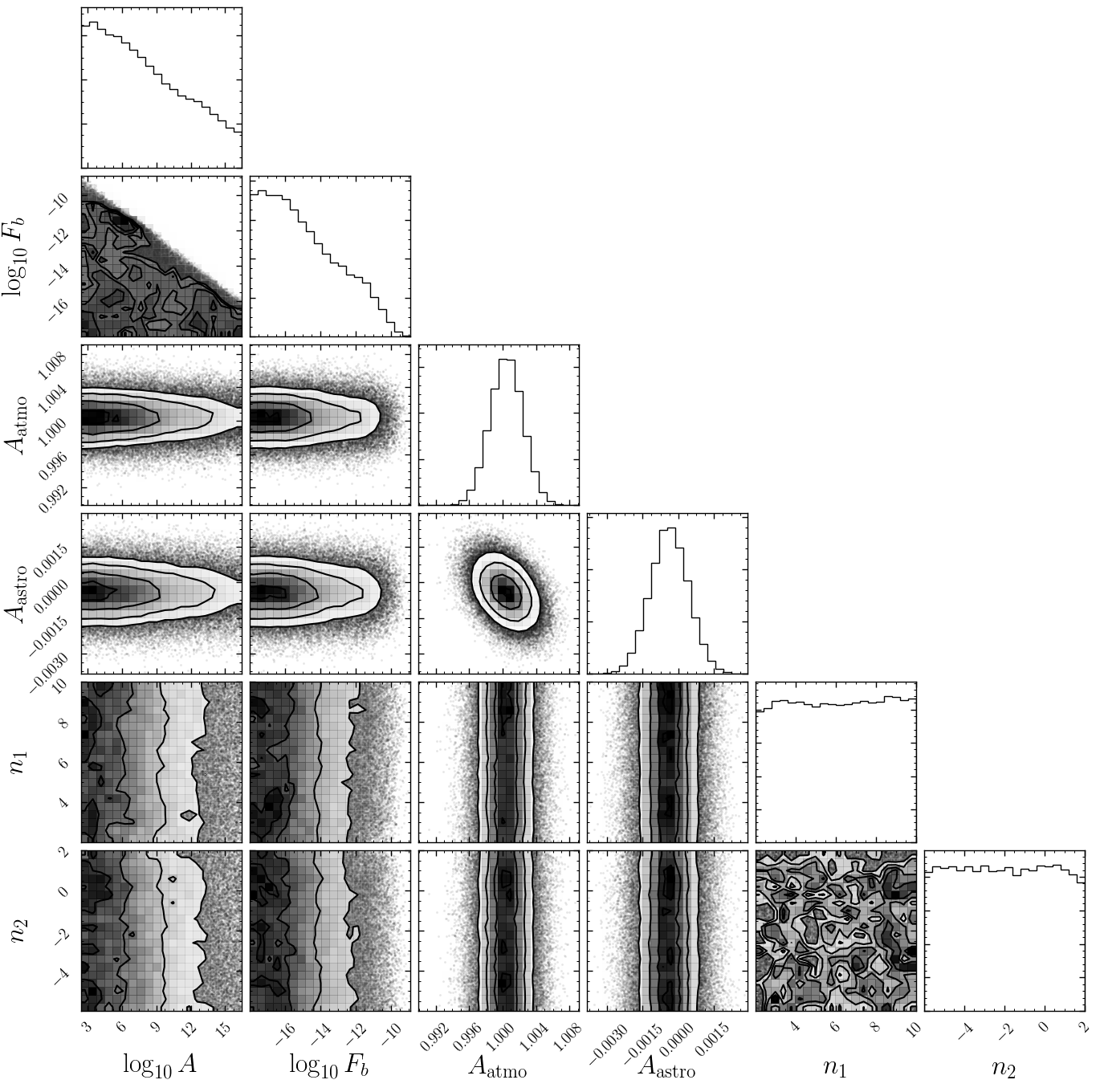}
\end{center}
\vspace{-0.5cm}
\caption{Triangle plot for the case of a point-source population following the SFD dust map.
The full posterior from which this was created is available for download.
$A$ is given in TeV$\,$cm$^2\,$s and $F_b$ in TeV$^{-1}\,$cm$^{-2}\,$s$^{-1}$.
}
\label{fig:results-corner}
\vspace{-0.4cm}
\end{figure*}

We can also consider the full posterior.
In Fig.~\ref{fig:results-corner} we show a triangle plot generated from the posterior for the case of the SFD dust signal template.
The signal parameters are clearly consistent with a background only hypothesis, and we note that the triangle plots for other templates are similar.
The posterior for each template is made publicly available, and we refer to App.~\ref{sec:PubPost} for details.

The purpose of the public posterior is that they can be used to test any point source population model where the associated $dN/dF$ can be approximated by a broken power law.
There are a wide number of source classes which have been considered as possible contributors to the IceCube neutrino flux.
For an overview, see, for example~\cite{Murase:2016gly}.
A fundamental problem, however, is that in many cases there remains considerable uncertainty in the associated luminosity function.
While we often have measurements of the photon luminosity function in the infrared, X-ray, or $\gamma$-ray energies, mapping from this to the neutrino luminosity function involves a number of assumptions.
For an example in the case of blazars, see~\cite{Yuan:2019ucv}.
For these reasons, the model space associated with neutrino sources is significant.

\begin{figure}[t]
\begin{center}
\includegraphics[scale=0.35]{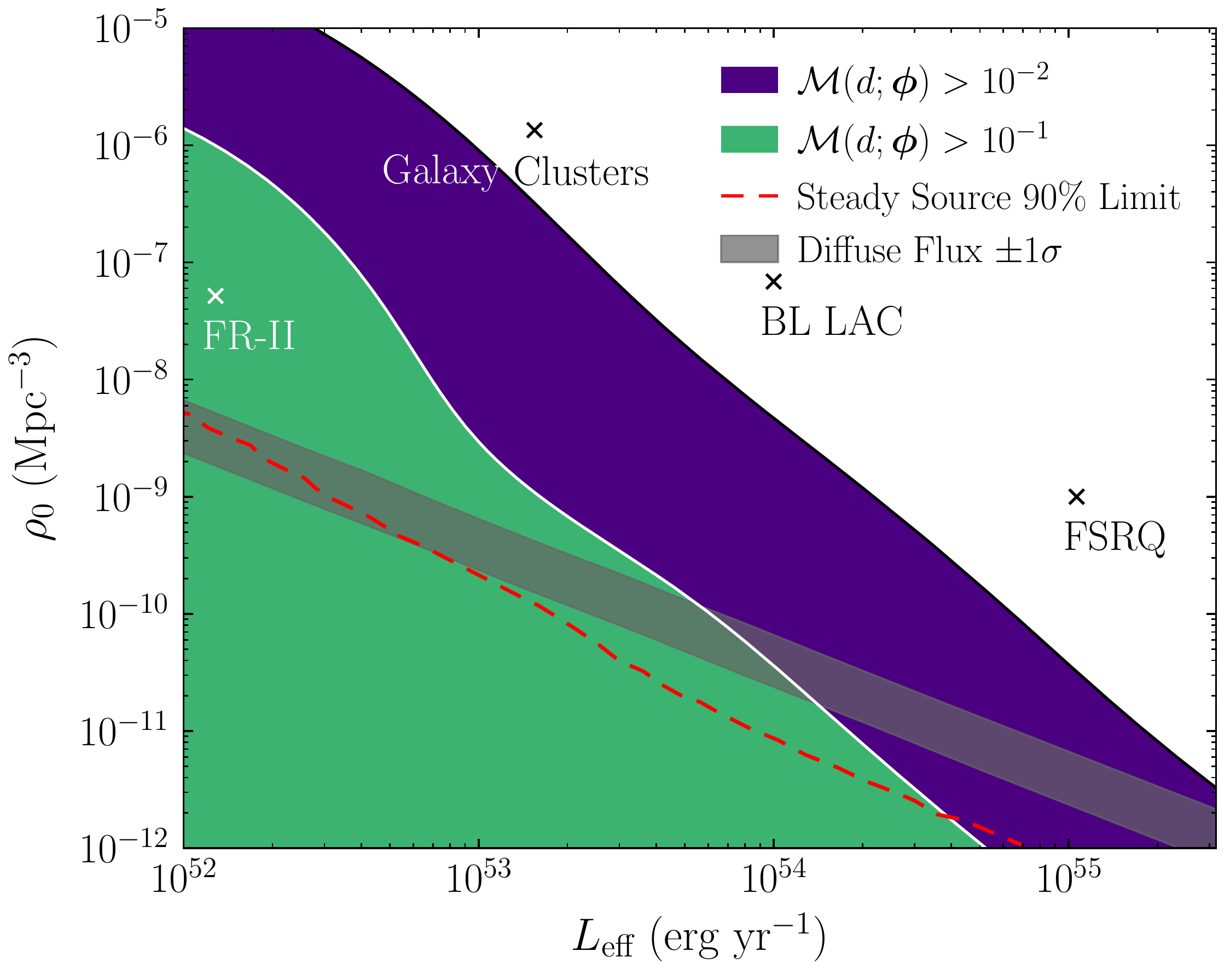}
\end{center}
\vspace{-0.5cm}
\caption{The pointwise likelihood ratio applied to the space of standard candle luminosity functions.
The standard candle approach models the luminosity function as sharply peaked in $L_{\rm eff}$, and then the space of possible models is spanned by this parameter and the density of sources, $\rho_0$.
This figure also allows us to contrast our results with the 90\% upper limit obtained using an analysis for steady point sources with a specified flux-distribution, as derived in~\cite{Aartsen:2018ywr}.
Both of these results can be compared to a standard candle population of sources that is compatible with the observed diffuse flux at $\pm 1 \sigma$, as quoted in~\cite{Aartsen:2017mau}.
To aid interpretation, we have overlaid the electromagnetic luminosities associated with several possible source classes: flat-spectrum radio quasars (FSRQ), BL Lacertae active galactic nuclei (BL LAC), galaxy cluster, and Fanaroff–Riley Class II radio galaxies (FR-II), following~\cite{Kowalski:2014zda,Taboada:2018gmk}.
We emphasize that these are not predicted neutrino luminosities, which are unknown, but highlight that current measurements provide information about the relative neutrino to photon luminosities of these sources.
We note that the results in this figure were derived using the NPTF posterior, described in App.~\ref{sec:PubPost}, and show the power of our result to test specific model hypotheses.
See text for details.
}
\label{fig:StdCandle}
\vspace{-0.4cm}
\end{figure}

One approach to simplifying this space is to consider standard candles.
Under this approach, the luminosity function is chosen to be sharply peaked at a certain value, denoted $L_{\rm eff}$, and then the problem is reduced to scanning a two dimension space parameterized by the effective luminosity, and the density of sources, denoted $\rho_0$.
More quantitatively, following~\cite{Aartsen:2018ywr}, the luminosity function of a standard candle is defined as a log-normal distribution with median $L_{\rm eff}$ and a width of 0.01 in $\log_{10} L_{\rm eff}$.
This model is then converted to an associated $dN/dF$ using \texttt{FIRESONG}~\cite{Taboada:2018gmk}, adopting a density evolution for the source population according to the evolution of the star formation rate in~\cite{Hopkins:2006bw}, and a flat universe with $\Omega_{M,0}=0.308$, $\Omega_{\lambda,0}=0.692$, and $h=0.678$~\cite{Ade:2015xua}.
These output source-count distributions were then interfaced with the NPTF posterior, and the value of $\mathcal{M}(d;\boldsymbol{\phi})$ calculated for each point in parameter space.
The $A$ and $F_b$ parameters scale with $\rho_0$ and $L_{\text{eff}}$ respectively, while $n_1$ and $n_2$ are set to $1.9$ and $-2$ respectively.
A lower limit of $L_{\text{eff}}=10^{52}$ erg yr${}^{-1}$ was chosen to match the prior on $F_b$.
The result is shown in Fig.~\ref{fig:StdCandle}, and -- consistent with our previous results -- we see no evidence for any particular source class.
These results, in addition to allowing a comparison with searches for source populations with fixed flux-characteristics~\cite{Aartsen:2018ywr}, also are representative of the power and generality of the NPTF technique.

%%%%%%%%%%%%%%%%%%%%%%%%%%%%%%%
\section{Conclusion}\label{sec:Conclusion}
%%%%%%%%%%%%%%%%%%%%%%%%%%%%%%%

In this work we have performed the first application of the non-Poissonian template fitting technique to search within the IceCube dataset for neutrino point-source populations.
Although IceCube presents novel challenges to the implementation of the NPTF, our work provides an explicit verification that such difficulties can be addressed, and that this technique is a viable method to search for such populations.
In addition to being able to search for populations with an, in principle, arbitrary source-count function $\mathrm{d}N/\mathrm{d}F$, this method also allows us to search for point sources with peculiar spatial distributions, and here we have considered spatial templates following maps of the isotropic sky, \textit{Fermi} bubbles, SFD dust map, and galactic disk.
In all cases, no significant evidence of a point-source population has been detected, and so we have presented limits in their absence, as shown in Fig.~\ref{fig:results}.
Importantly, we have made the full posterior from our analysis publicly available, allowing specific theory predictions for contributions to the IceCube flux to be tested directly.
This is exemplified by the application of our results to the space of standard candle luminosity functions, shown in Fig~\ref{fig:StdCandle}.

There are a number of ways that the analysis presented here can be improved upon.
The IceCube dataset contains a large amount of information on the reconstruction quality of incident candidate neutrinos on an event by event basis.
As the NPTF is a fundamentally binned method, much of this information is lost, and is only exploited through the optimization of various high level cuts, such as on the energy range considered.
Yet there is significant scope to incorporate more of this information into the NPTF.
For example, there is the potential to incorporate energy binning into the method, and with this additional event information.
Beyond expanding the neutrino dataset, such extensions could play an important role in uncovering evidence for a population of astrophysical point sources, and unravelling the mystery surrounding the origin of the IceCube neutrinos.

%%%%%%%%%%%%%%%%%%%%%%%%%%%%%%%%%%%%%%%%%%
\begin{acknowledgments}
%%%%%%%%%%%%%%%%%%%%%%%%%%%%%%%%%%%%%%%%%%

The IceCube Collaborations acknowledges the significant contributions to this manuscript from Gabriel Collin, Jimmy DeLaunay, and Nicholas Rodd.
The authors gratefully acknowledge the support from the following agencies and institutions:
USA {\textendash} U.S. National Science Foundation-Office of Polar Programs,
U.S. National Science Foundation-Physics Division,
Wisconsin Alumni Research Foundation,
Center for High Throughput Computing (CHTC) at the University of Wisconsin-Madison,
Open Science Grid (OSG),
Extreme Science and Engineering Discovery Environment (XSEDE),
U.S. Department of Energy-National Energy Research Scientific Computing Center,
Particle astrophysics research computing center at the University of Maryland,
Institute for Cyber-Enabled Research at Michigan State University,
and Astroparticle physics computational facility at Marquette University;
Belgium {\textendash} Funds for Scientific Research (FRS-FNRS and FWO),
FWO Odysseus and Big Science programmes,
and Belgian Federal Science Policy Office (Belspo);
Germany {\textendash} Bundesministerium f{\"u}r Bildung und Forschung (BMBF),
Deutsche Forschungsgemeinschaft (DFG),
Helmholtz Alliance for Astroparticle Physics (HAP),
Initiative and Networking Fund of the Helmholtz Association,
Deutsches Elektronen Synchrotron (DESY),
and High Performance Computing cluster of the RWTH Aachen;
Sweden {\textendash} Swedish Research Council,
Swedish Polar Research Secretariat,
Swedish National Infrastructure for Computing (SNIC),
and Knut and Alice Wallenberg Foundation;
Australia {\textendash} Australian Research Council;
Canada {\textendash} Natural Sciences and Engineering Research Council of Canada,
Calcul Qu{\'e}bec, Compute Ontario, Canada Foundation for Innovation, WestGrid, and Compute Canada;
Denmark {\textendash} Villum Fonden, Danish National Research Foundation (DNRF), Carlsberg Foundation;
New Zealand {\textendash} Marsden Fund;
Japan {\textendash} Japan Society for Promotion of Science (JSPS)
and Institute for Global Prominent Research (IGPR) of Chiba University;
Korea {\textendash} National Research Foundation of Korea (NRF);
Switzerland {\textendash} Swiss National Science Foundation (SNSF);
United Kingdom {\textendash} Department of Physics, University of Oxford.
We thank Derek Fox and Kohta Murase for useful discussions.
The work presented in this paper made use of \texttt{astropy}~\cite{2013A&A...558A..33A}, \texttt{corner}~\cite{corner}, \texttt{FIRESONG}~\cite{Taboada:2018gmk}, \texttt{HEALPix}~\cite{Gorski:2004by}, \texttt{MultiNest}~\cite{Feroz:2008xx,Buchner:2014nha}, and \texttt{NPTFit}~\cite{Mishra-Sharma:2016gis}.
NLR acknowledges support from the Miller Institute for Basic Research in Science at the University of California, Berkeley.

%%%%%%%%%%%%%%%%%%%%%%%%%%%%%%%%%%%%%%%%%%
\end{acknowledgments}
%%%%%%%%%%%%%%%%%%%%%%%%%%%%%%%%%%%%%%%%%%

\newpage
\appendix

%%%%%%%%%%%%%%%%%%%%%%%%%%%%%%%
\section{Description of the Public Posterior}\label{sec:PubPost}
%%%%%%%%%%%%%%%%%%%%%%%%%%%%%%%

\newcommand{\code}[1]{\texttt{\detokenize{#1}}}

The posterior for each of the four templates can be found at \url{https://icecube.wisc.edu/science/data/NPTF_7yr_posterior} as an HDF5 file.
Within the file, five tables named \code{Isotropic}, \code{Galactic_disk}, \code{Fermi_bubble}, \code{SFD_dust}, and \code{Northern_sky} contain the posterior for their respective templates.

Each table describes equally-weighted samples using five columns.
Four columns, labelled \code{ln_A},  \code{ln_Fb}, \code{n1}, and \code{n2} contain the coordinates for the sample in natural logarithmic parameter space for the differential source-count function normalization $A$ and break $F_b$, while the power indices $n_1$ and $n_2$ are in linear space.
The fifth column -- labelled \code{loglikelihood} -- gives the natural logarithm of the likelihood function at the location of the corresponding sample.
In addition, each table has two attributes named \code{P_log_evidence} and \code{NP_log_evidence} that contain the natural logarithm of the evidence integral for the Poissonian model ($\mathcal{L}_0$) and non-Poissonian model ($\mathcal{L}_1$) respectively.

The root node of the HDF5 file also contains a series of attributes named \code{units_ln_A}, \code{units_ln_Fb}, \code{units_n1}, and \code{units_n2} that specify the units that the posterior sample coordinates are given in.
Another series of root attributes named \code{prior_ln_A}, \code{prior_ln_Fb}, \code{prior_n1}, and \code{prior_n2} give the probability density of the uniform priors for each of the model parameters.

Finally, we emphasize that our analysis and hence these posteriors are constructed with the assumption that the astrophysical population produces neutrinos with an $E^{-2}$ spectrum, as given in \eqref{eq:PhiE}.

\begin{figure*}[t]
\begin{center}
\includegraphics[scale=0.35]{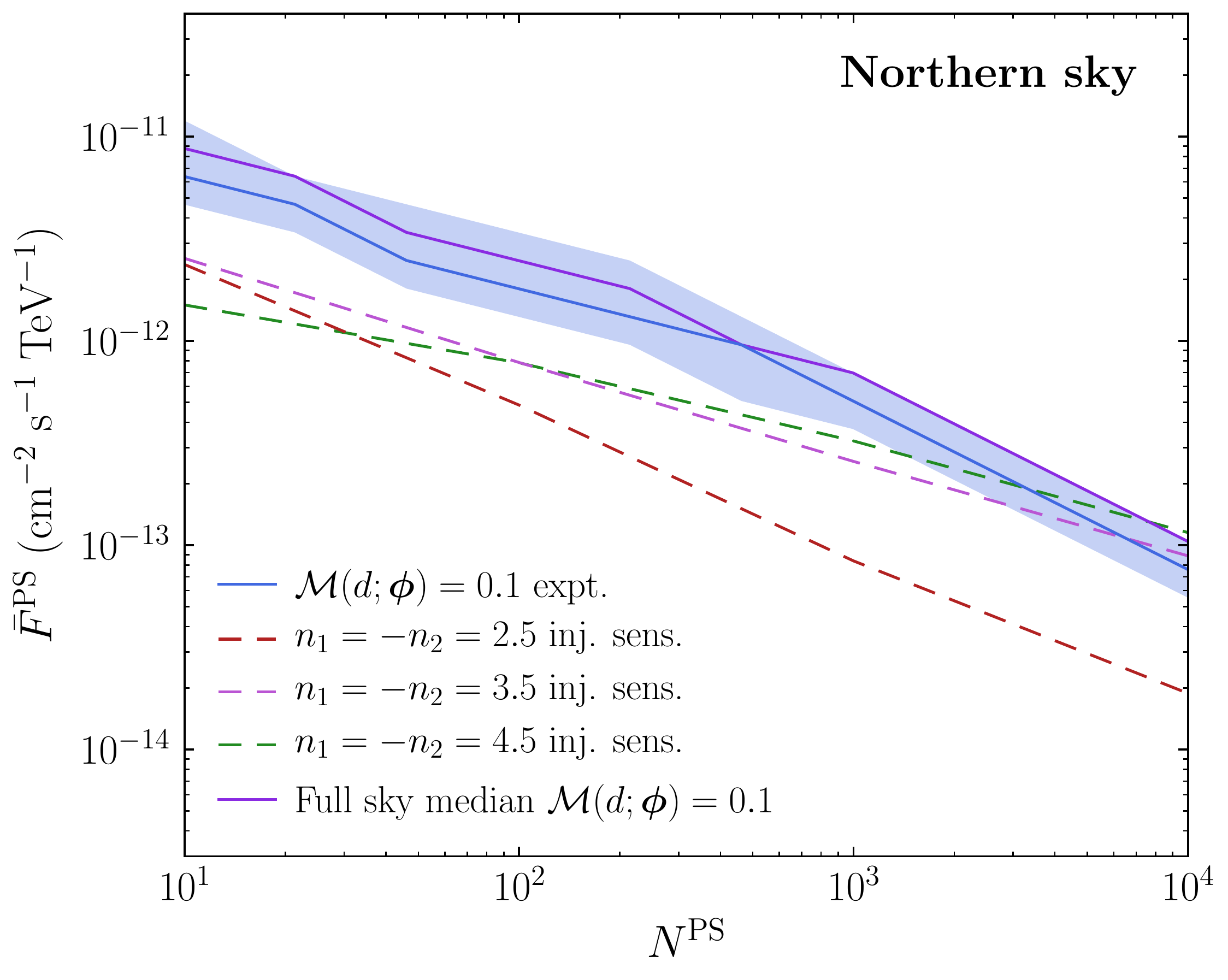}
\hspace{0.5cm}
\includegraphics[scale=0.355]{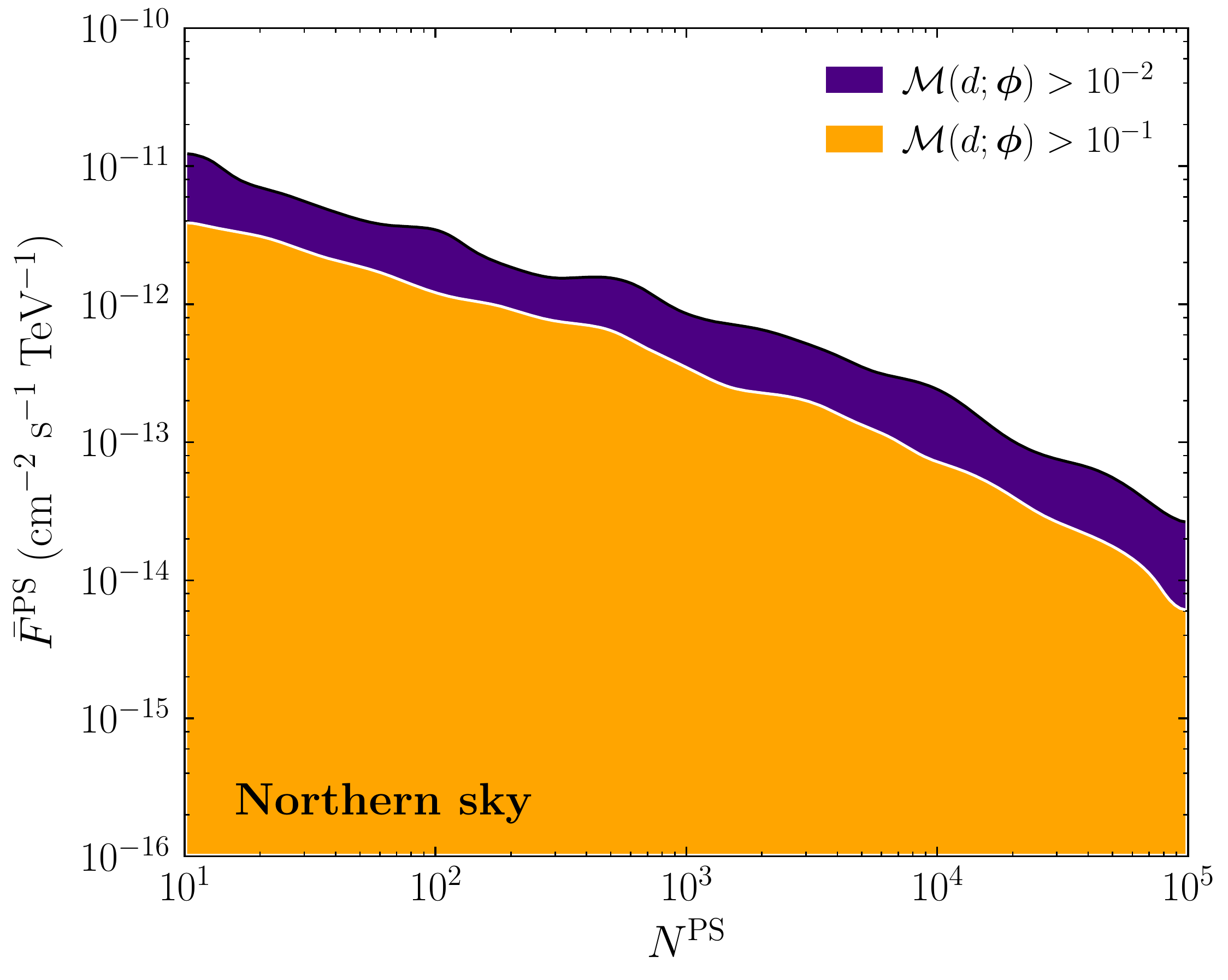}
\end{center}
\vspace{-0.5cm}
\caption{The analogue of the full sky sensitivity shown in Fig.~\ref{fig:sensitivity} (top) and the pointwise likelihood ratio as in Fig.~\ref{fig:results} (bottom), but here restricted only to the northern hemisphere.
Interestingly we see at most a factor of a few improvement in reach, suggesting that the NPTF technique is not being hampered by the increased background in the southern hemisphere.
For the top figure we have explicitly reproduced the median of the full sky $\mathcal{M}(d; {\boldsymbol \phi})=0.1$ distribution to allow a direct comparison.
}
\label{fig:northfullsky}
\vspace{-0.4cm}
\end{figure*}

%%%%%%%%%%%%%%%%%%%%%%%%%%%%%%%
\section{Isotropic Sources in the Northern Sky}\label{sec:NorthIso}
%%%%%%%%%%%%%%%%%%%%%%%%%%%%%%%

Traditional searches for extragalactic point sources at IceCube are performed restricting to the northern or southern hemispheres.
The motivation for this is the northern hemisphere, having a lower background, usually has an enhanced sensitivity.
In Fig.~\ref{fig:northfullsky} we show the expected sensitivity and the pointwise likelihood ratio determined from the data for the northern sky, which should be contrasted to the full sky result for both hemispheres shown in Figs.~\ref{fig:sensitivity} and~\ref{fig:results}.
Comparing the two results, it is clear that restricting the NPTF to the lower background hemisphere only marginally improves the sensitivity.
This suggests that the NPTF results are not being degraded by working with the full sky, and given our inclusion of a number of galactic templates, justifies the choice of both hemispheres used in the main text.
\newpage

\bibliography{IceCube-NPTF}

\end{document}